\documentclass[11pt,a4paper]{article}
\usepackage[round]{natbib} 
\usepackage{amsmath} 
\numberwithin{equation}{section} 
\usepackage{graphicx} 
\usepackage{here} 
\usepackage{enumitem} 
\usepackage[font=footnotesize]{caption} 
\usepackage{hyperref}
\usepackage[hypcap]{caption}

\usepackage{Sweave}
\begin{document}
\title{How to speed up R code: an introduction}
\author{
    Nathan Uyttendaele\footnote{Universit\'e catholique de Louvain, Institut de Statistique, Biostatistique et Sciences Actuarielles, Voie du Roman Pays 20, B-1348 Louvain-la-Neuve, Belgium.} \\
    \small na.uytten@gmail.com
}
\date{\today}
\maketitle

\begin{abstract}
Most calculations performed by the average R user are unremarkable in the sense that nowadays,
any computer can crush the related code in a matter of seconds. But more
and more often, heavy calculations are also performed using R, something especially true in some fields such as statistics. The user then faces total execution times
of his codes that are hard to work with: hours, days, even weeks. In this paper, how to
reduce the total execution time of various codes will be shown and typical bottlenecks
will be discussed. As a last resort, how to run your code on a cluster of computers (most
workplaces have one) in order to make use of a larger processing power than the one
available on an average computer will also be discussed through two examples.
\end{abstract}

\newpage

\tableofcontents
\newpage
\section{Your computer}

So you are suffering from a slow R code. How to speed up that code? 

A very first (naive) thing you could try is to buy a better computer than the one you have been using to run your code, especially if that computer is more than 10 years old. While the fact a same R function will usually run faster on a brand new computer bought at the local store is not a surprise, it is nonetheless important to understand why.

There are 3 hardware parts of your computer capable of impacting the speed of an R function (or more generally of your R code):
\begin{itemize}[noitemsep, nolistsep]
\item the hard disk drive (HDD)
\item the random-access memory (RAM)
\item the central processing unit (CPU)
\end{itemize}

In the following subsections, each of these hardware parts is briefly discussed.

\subsection{HHD}
The hard disk drive is where all the data of the computer is stored, even when it is powered down. Your personal files such as pictures or musics are stored there. If you have large datasets you want to analyze in R, those datasets will also be stored there. R itself is stored there! 

Because they are very cheap to produce, hard disk drives are the best way to store large amount of data. But unfortunately, they are slow. 

To load a 4-gigabyte dataset (which by the way will require you to use a 64-bit version of R, otherwise you will get an error, more on that later), you issue an R command such as

\begin{Verbatim}[frame=single, samepage=true]
data=read.table("data.txt")
\end{Verbatim}

And because the hard disk drive is slow and the dataset is huge, you will have to wait.

The same problem occurs whenever you want to write down your results. To save a 4-gigabyte dataset after some changes, you issue the command

\begin{Verbatim}[frame=single, samepage=true]
write.table(data, "data.txt")
\end{Verbatim}

This will erase the file \verb|data.txt|. Again you will have to wait, as the process of writing down the \verb|data| object will take some time.

How long are we talking? Try this yourself. Submit the command
\begin{Verbatim}[frame=single, samepage=true]
data=rnorm(10^7)
\end{Verbatim}
to create a large vector, and now write it down on the hard disk drive with the command

\begin{Verbatim}[frame=single, samepage=true]
write.table(data, "data.txt")
\end{Verbatim}

The computer used at the time of writing required about 40 seconds to perform this last task. The weight of the resulting file \verb|data.txt| was 280.5 MBs. It can therefore be roughly inferred that a 4-GB dataset will require 500 seconds to be written down on a hard disk drive such as the one used at the time of this little experiment.

Knowing this, a very first rule when writing R code is the following: only access the hard disk drive at the beginning of your code (to load some data) and at the end (to save some results). Reading or writing down data over and over again while in the main body of your code can be very expensive and should be avoided.

Continuous improvements in the speed of hard disk drives have been made over the years, mainly by making them spinning faster and faster. A few years ago, there was even a breakthrough: a new type of drive called SSD for Solid-State Drive was introduced. These new drives do not require any physical part of the drive to move anymore and are faster than usual hard disk drives. They still remain one of the slowest parts of any computer. SSDs are however less easy to break, smaller, consume less energy and are therefore quite often encountered in laptops. As an important drawback, they are more expensive to produce. This has lead to a generation of computers having both a solid-state drive and a hard disk drive: you store files you have to access quite often on the small SSD and files of lesser interest on the large HDD. On October 23, 2012, Apple introduced the Fusion Drive, which is basically a SSD and a HDD put together in a way that the user cannot distinguish them anymore: the operating system automatically manages the content of the drive so that the most frequently accessed files, applications, documents, photos and other data are stored on the faster solid-state drive, while infrequently used items are automatically moved on the hard disk drive. A very good idea, although there are no doubts that hard disk drives will eventually disappear and that all computers will eventually end up equipped with SSDs only. As of 2015, it is getting almost impossible to find a laptop equipped with a HDD in usual stores in Belgium: SSDs are already the norm among brand new laptops (HHDs however remain common on desktop computers).

\subsection{RAM}

Like the hard disk drive, the random-access memory is also used to store files. What happens when you load a dataset stored on the hard disk drive by using \verb|read.table|? The dataset is copied from the HDD (or SSD) and pasted on the RAM. This takes some time: remember, accessing the hard disk drive is expensive. Any further command you submit in order to interact with the dataset will however be executed almost instantly. Why? Because the RAM is a much faster data storage technology, giving you the ability to interact with the dataset in real-time or at least much faster than if it was still stored on the hard disk drive only. In layman's terms: the dataset is made ``hot" and you will not have to wait anymore to interact with it. 

When you open R itself or any other program on your computer, it also takes some time for exactly the same reason: the operating system is making the program ``hot", meaning it is copied from the (slow) hard disk drive toward the (fast) RAM. This way you can interact with the program without having to wait each time you perform an action. The term ``loading" usually refers to the process of copying a file from the hard disk drive toward the random-access memory.

Why are we still making hard disk drives or solid-state drives while obviously we want to store all of our files on the RAM?

Unlike a hard disk drive, everything stored on the RAM is lost when the computer is powered down. This is one problem. The other reason is that RAM is a very expensive data storage technology: replacing a 512-GB hard disk drive by 512 GBs of random-access memory will lead to a computer with an unacceptable price tag.

Why does RAM matter for R? First, RAM produced ten years ago is not as fast as RAM produced nowadays. Second, R will suffer greatly, as for any other program, if you run out of random access memory. Loading a 10-gigabyte dataset from the hard disk drive to the RAM while there are only 8 gigabytes of RAM available will result in a very slow code. Indeed, when you run out of memory, the computer has no choice but to use part of the hard disk drive as if it was random-access memory. Considering how expensive it is to access the hard disk drive, it is easy to imagine how ugly things can become in that scenario.

In the HDD subsection was mentioned that, unless you are using 64-bit R, you will not be able to load a 4-GB dataset. Old computers do not have a lot of RAM and run 32-bit programs. Brand new computers have more RAM and run both 64-bit programs and 32-bit programs. 32-bit programs cannot use more than 2 GBs of RAM. 64-bit programs are allowed to use as much RAM as they need (unless of course you run out of memory).

Do you run 64-bit R or 32-bit R? If you have no idea, simply open a fresh R instance and read the first lines in the console:

\begin{Verbatim}[frame=single, samepage=true]
R version 3.1.1 (2014-07-10) -- "Sock it to Me"
Copyright (C) 2014 The R Foundation for Statistical Computing
Platform: x86_64-w64-mingw32/x64 (64-bit)
\end{Verbatim}

\verb|Platform: x86_64-w64-mingw32/x64 (64-bit)| points to a 64-bit version of R.

Note: as years will pass, we can expect 32-bit programs to disappear and 32-bit R to be available only for compatibility purposes. 

\subsection{CPU}

Last but not least, the central processing unit. This is the heart of your computer, where all calculations are done. When you ask R to add two numbers, the CPU makes this addition for R. If you ask R to perform $10^9$ additions, the CPU is put under more pressure. The faster the CPU, the faster the calculations are done. 

Over the years the speed of CPUs has improved at an amazing rate (see Moore's law). However at some point and for physical reasons, it was almost impossible to further improve the speed of CPUs. So engineers had an amazing idea: why not put more than one CPU in every computer? As a result, we now live in a ``multi-core" world, where most of our devices have several ``cores" (CPUs). It is said that such devices possess a multi-core processor. Wikipedia gives us the following definition:

\begin{quote}
A multi-core processor is a single computing component with two or more independent actual central processing units (called cores), which are the units that read and execute program instructions.
\end{quote}

Unfortunately, R is not natively able to use several cores at the same time! This is true for most other programs as well. If you use a computer with 8 cores (8 CPUs) and ask R to perform $10^9$ additions, do not expect 1/8 to be done on the first core, 1/8 on the second core, etc. All calculations will be done using one and only one core at the same time, meaning 12.5\% of the total computing power available will be used by R at best. The other cores will remain unused. Later in this paper, some packages allowing R to use more than one core at the same time will be explored. The strategy of opening R several times and of breaking down the calculations across these different R instances in order to use more than one core at the same time will also be explored (this strategy is very effective!)

\subsection{Monitoring your computer}

Monitoring how hard the cores of a computer are working or the amount of RAM used in real-time is possible. Hereafter is described how it works for Windows, Mac OS and Linux Ubuntu.

First, Windows 7. Press Ctrl + Shift + Esc (works for Windows 8 and presumably for Windows 10, too). Something called the Windows Task Manager will open. Pick the Performance tab. 

\begin{figure}[H]

\centering
\includegraphics[width=0.8\textwidth]{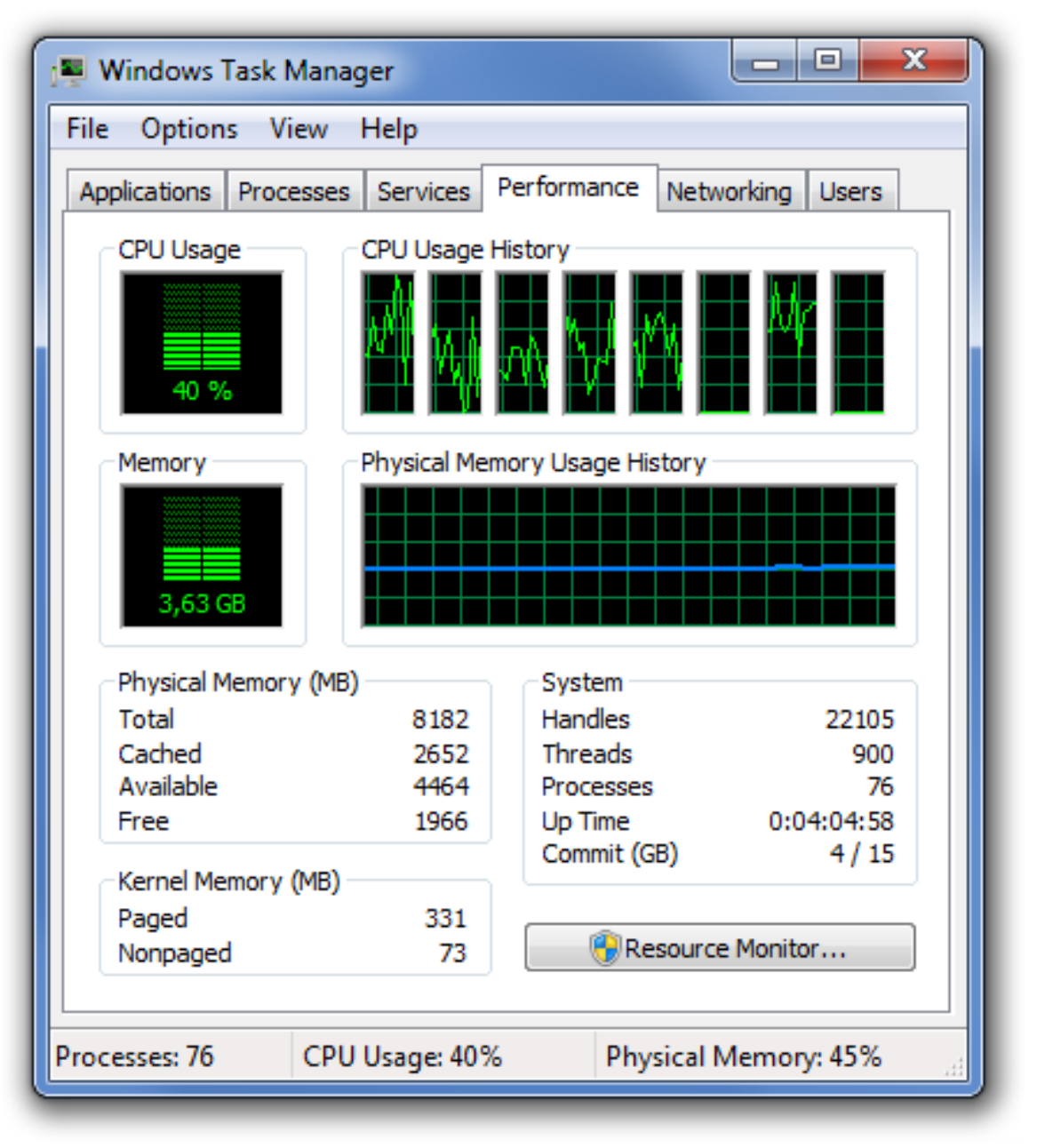} \\
\caption{The Windows Task Manager, Performance tab. \label{windows_task_manager}}
\end{figure}

CPU Usage gives you in real-time how much of your multi-core processor is used (Figure \ref{windows_task_manager}). The number of rectangles below CPU Usage History is equal to the number of cores you have in your multi-core processor. It is also possible to check the cores that are working and the cores that are not. Two cores among 8 were doing nothing at the time of capture.

Memory simply gives how much RAM is used in real-time. 3.63 GB of RAM among 8 GB was used at the time of capture.

\begin{figure}[H]
\centering
\includegraphics[width=0.9\textwidth]{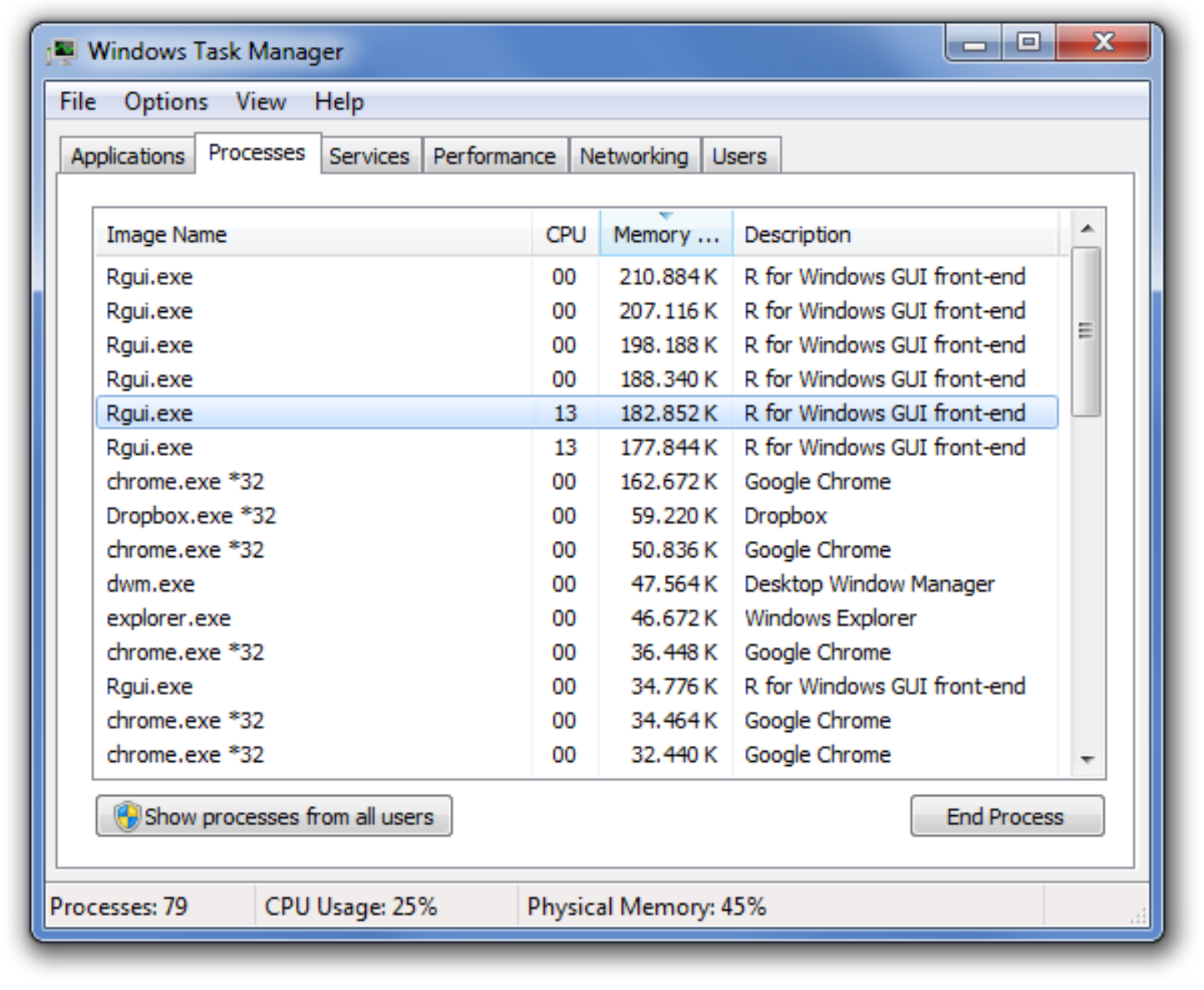} \\
\caption{The Windows Task Manager, Processes tab. \label{windows_task_manager2}}
\end{figure}

Another interesting tab is the Processes tab. In here, all the programs currently active are listed (Figure \ref{windows_task_manager2}). In this paper, an active program or process is defined as a program that was loaded from the hard disk drive toward the RAM, enabling the user to interact with that program in real-time. The RAM an active program or process uses varies over time, as well as the amount of computing power. The amount of computing power used by an active process is usually near 0 when that process has nothing to do and is waiting for some event, such as user input. Keep in mind a same program can usually be made active (we can also say opened) several times.

The column CPU in Figure \ref{windows_task_manager2} gives you how much of the \textit{total} computing power is currently used by the corresponding process. Several R instances were active at the time of this second capture (not the same time as the previous capture) and only two of these R instances were actually working hard, using each a different core, meaning 25\% of the total computing power available was being used. The 4 other R instances were waiting for user input at the time of capture. The memory usage in KB for each program is also available in the ``Memory...'' column.

\begin{figure}[H]

\centering
\includegraphics[width=0.75\textwidth]{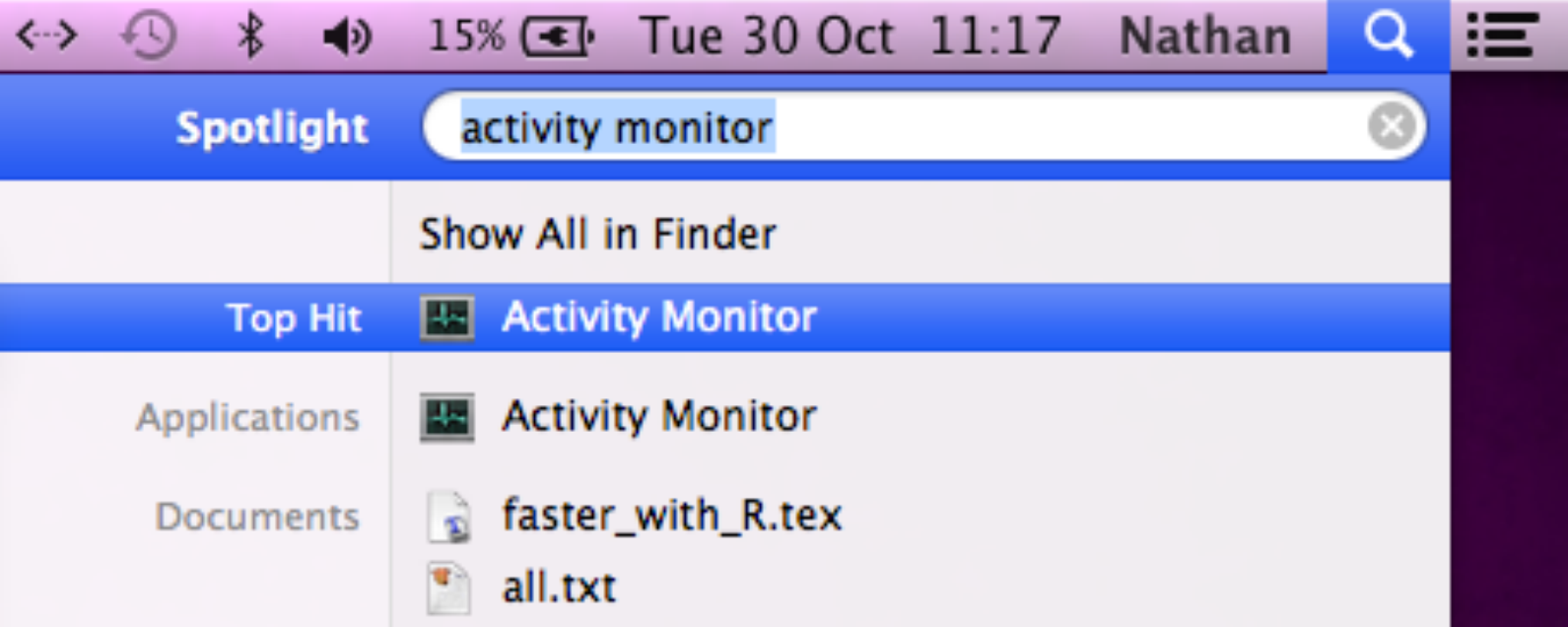} \\
\caption{spotlight search for Activity Monitor \label{spotlight}}
\end{figure}

Second, Mac OS. On Mac OS, what you want to look for is called ``Activity Monitor". Use spotlight, as shown in Figure \ref{spotlight}.

\begin{figure}[H]

\centering
\includegraphics[width=0.9\textwidth]{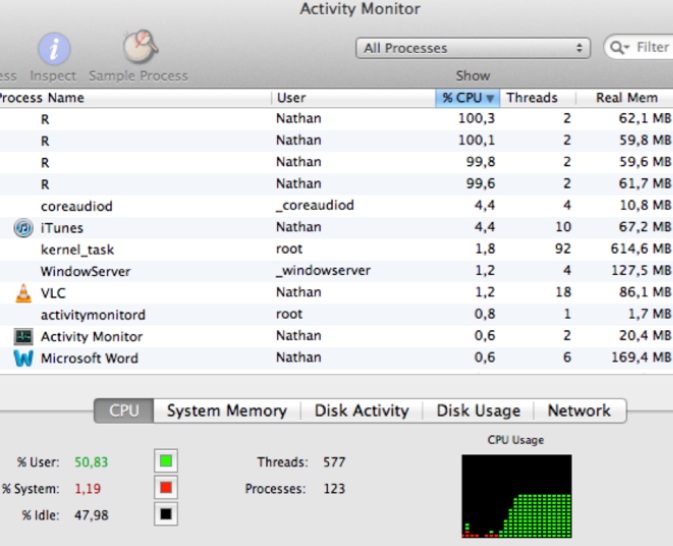} \\
\caption{Activity Monitor. \label{activity_monitor}}
\end{figure}

Activity Monitor gives you a list of all active programs on the computer. In this paper, an active program or process is defined as a program that was loaded from the hard disk drive toward the RAM, enabling the user to interact with that program in real-time. The RAM an active program or process uses varies over time, as well as the amount of computing power. The amount of computing power used by an active process is usually near 0 when that process has nothing to do and is waiting for some event, such as user input. Keep in mind a same program can usually be made active (we can also say opened) several times. As seen in Figure \ref{activity_monitor}, four R instances were opened, each R instance working hard and using about 100\% of the core assigned to it. Real Mem gives you the memory usage for each program: for instance, iTunes was using 67.2 MB of memory at the time of capture.

The number of cores you have can be inferred by having several programs working hard at the same time and by looking at CPU Usage, bottom right of Figure \ref{activity_monitor}. With 4 R instances working hard, one can see that half of the total computing power available was used. Therefore, it can be inferred that the total number of cores was 8 for the Mac used at the time of capture.

\begin{figure}[H]

\centering
\includegraphics[width=0.9\textwidth]{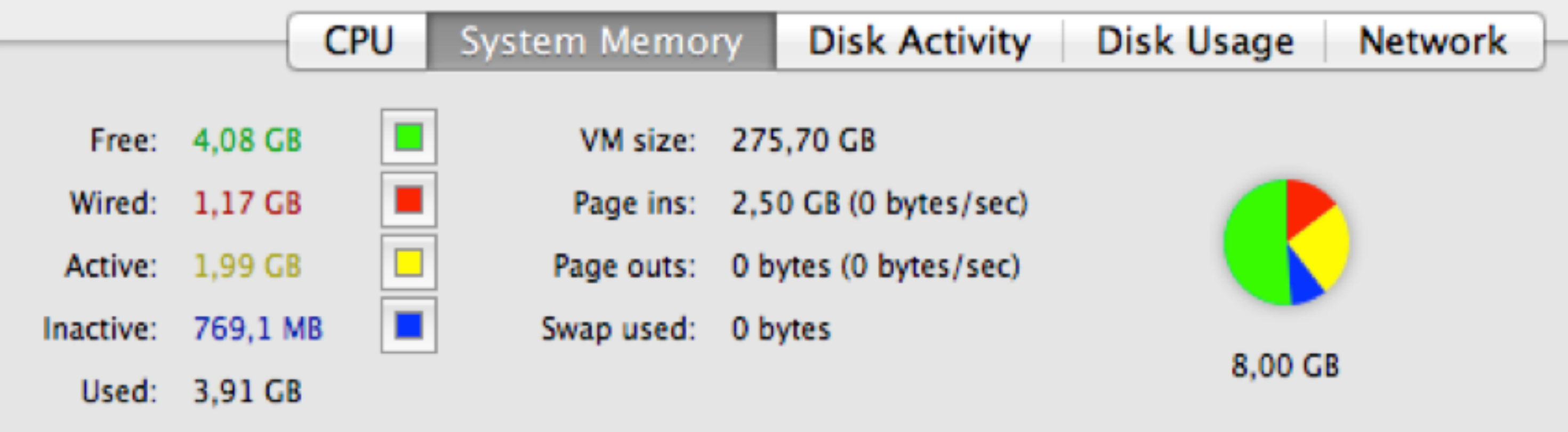} \\
\caption{System Memory.\label{activity_monitor2}}
\end{figure}

If you click on System Memory, a pie chart of the total memory usage appears. Figure \ref{activity_monitor2} shows that about half of the 8 GBs of memory was used and the other half was free at the time of capture.

\begin{figure}[H]
\centering
\includegraphics[width=0.45\textwidth]{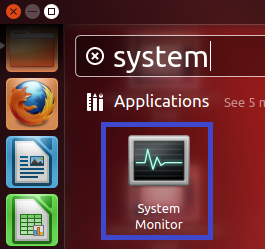} \\
\caption{\label{dash}}
\end{figure}

Last, Linux Ubuntu. Use the Dash to find something called ``System Monitor", as shown in Figure \ref{dash}.

The Processes tab in System Monitor lists all programs currently active on the related computer. In this paper, an active program or process is defined as a program that was loaded from the hard disk drive toward the RAM, enabling the user to interact with that program in real-time. The RAM an active program or process uses varies over time, as well as the amount of computing power. The amount of computing power used by an active process is usually near 0 when that process has nothing to do and is waiting for some event, such as user input. Keep in mind a same program can usually be made active (we can also say opened) several times.

\begin{figure}[H]

\centering
\includegraphics[width=0.8\textwidth]{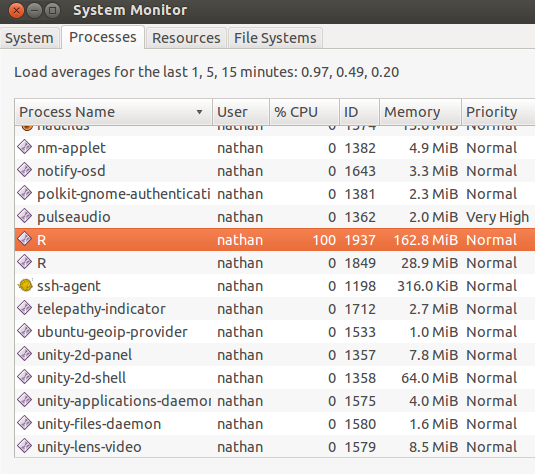} \\
\caption{System Monitor, Processes tab. \label{system_monitor}}
\end{figure}

Figure~\ref{system_monitor} shows that two R instances were active at the time of capture, one using its assigned core at 100\%, the other one doing nothing (not using its assigned core). The column Memory gives you the memory usage for each program in MiB (roughly equal to MB, see Wikipedia for more information about MiB vs MB, KiB vs KB, etc).

\begin{figure}[H]

\centering
\includegraphics[width=0.85\textwidth]{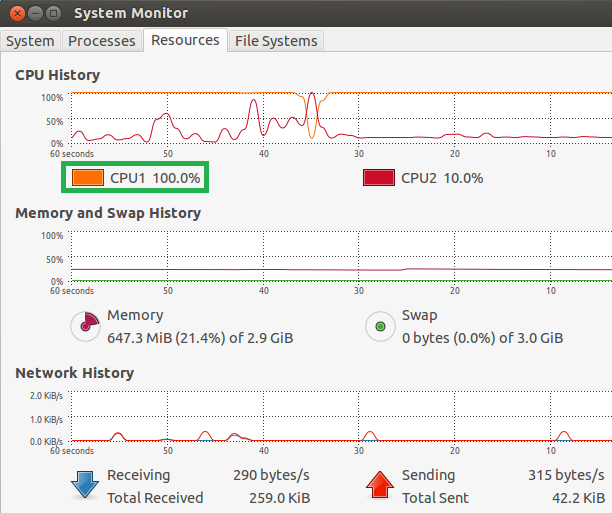} \\
\caption{System Monitor, Resources tab. \label{system_monitor2}}
\end{figure}

The Resources tab shows how many cores the computer is equipped with: this number of cores is equal to the number of different colors displayed below CPU History. Figure \ref{system_monitor2} shows there are two cores (one orange, one red) and the orange core was being used at 100\% at the time of capture.

The total amount of memory used by the computer (Memory and Swap History) was 647.3 MiBs at the time of capture, that is, 21.4\% of the total amount of memory available. Swap memory is a part of the hard disk drive intended to be used in case you run out of memory. Should the 2.9 GiBs of RAM and the 3.0 GiBs of Swap memory be used, things will go bad: the computer freezes and needs a hard reboot (this is called overswapping).

\section{One R instance, one core}
As mentioned in the previous section, R only uses one core at the time, even if there are more available. Various solutions to this problem will be given in the next section. In this section, how to make the best of the one core assigned to a given R instance is explored.

First you need to be able to measure how fast your code currently run. There are two ways to do this:

\begin{Verbatim}[frame=single, samepage=true]
time.start=proc.time()[[3]]

### Your code ###

time.end=proc.time()[[3]]
time.end-time.start
\end{Verbatim}

The result is given in seconds. The other way to measure the execution time is

\begin{Verbatim}[frame=single, samepage=true]
system.time({

### Your code ###

})[[3]]
\end{Verbatim}

The result is also given in seconds. Both ways will give the same result.

Example:

\begin{Verbatim}[frame=single, samepage=true]
> system.time({
+ data=rnorm(10^7)
+ })[[3]]
[1] 1.044
> system.time({
+ data=rnorm(10^7)
+ })[[3]]
[1] 1.005
>|
\end{Verbatim}

It is important to notice that the execution times you measure are, in general, random. If you run exactly the same code several times, expect to get different measured execution times. 

When comparing two different codes or functions leading to the same result in order to pick the fastest, the user should therefore measure the execution time of the first function $n_1$ times and the execution time of the second function $n_2$ times before performing a t-test (or a nonparametric test in case the assumption of normality is doubtful) on the measured execution times in order to decide which function is the fastest.

Fortunately, measuring the execution time of the first and second function only once is usually more than enough to determine with a high level of certainty the fastest function without having to use any statistical tool.

Along with the ability of measuring how fast your code run, it is important to be able to master the memory usage of an R instance. 

Any object created in R is stored in the RAM. As an example, if $10^7$ observations from a normal distribution are requested issuing
\begin{Verbatim}[frame=single, samepage=true]
data=rnorm(10^7)
\end{Verbatim}
about 80 MBs of RAM will be locked to store the $10^7$ values. What if the vector of values is not needed anymore? How can the 80 MBs of RAM be freed? Closing the related R instance is one solution (everything created inside that R instance will be unloaded from the memory). Another one is to issue the following commands:

\begin{Verbatim}[frame=single, samepage=true]
rm(data)
gc()
\end{Verbatim}

As an exercise, the reader is invited to perform the following experiment. While monitoring your R instance using Windows Task Manager, Activity Monitor or System Monitor, submit the command \verb|data=rnorm(10^7)| and observe the memory usage for the related R instance go up 80 MBs. The memory usage will go down 80 MBs when the commands \verb|rm(data)| and \verb|gc()| are submitted.

Submitting the command 
\begin{Verbatim}[frame=single, samepage=true]
ls()
\end{Verbatim}
will return the names, as a character vector, of all objects previously created in the related R instance.

Example:
\begin{Verbatim}[frame=single, samepage=true]
> ls()
[1] "EBR"         "SampleSize"  "SimuData"    "TrueBeta"
[5] "TrueTheta"      "V"           "j"           "mframe" 
[9] "to_profile5"
\end{Verbatim}

To remove several objects at once, issue

\begin{Verbatim}[frame=single, samepage=true]
rm(mframe, to_profile5, EBR)
gc()
\end{Verbatim}

or

\begin{Verbatim}[frame=single, samepage=true]
rm(list=c("mframe", "to_profile5", "EBR"))
gc()
\end{Verbatim}

To remove all objects from the RAM, issue the following command:

\begin{Verbatim}[frame=single, samepage=true]
rm(list=ls())
gc()
\end{Verbatim}

Note: these commands are expensive. They should not be used over and over in your code if it is not needed.

\subsection{Writing your code right}

Now that execution times can be measured, it is time for a first case study:

\begin{Verbatim}[frame=single, samepage=true]
a=NULL
for(i in 1:n) a=c(a,i^2)
\end{Verbatim}

This code creates a vector \verb|a| of size \verb|n| containing the squared integers, that is, $1^2$, $2^2$, \ldots, $\verb|n|^2$. In this code, \verb|a| is first declared as a vector of unknown size with

\begin{Verbatim}[frame=single, samepage=true]
a=NULL
\end{Verbatim}
and then is grown element by element, one more at each iteration of the for loop.

Another code leading to exactly the same result is

\begin{Verbatim}[frame=single, samepage=true]
a=vector(length=n)
for(i in 1:n) a[i]=i^2
\end{Verbatim}
but this time the vector \verb|a| is declared to its final size before the loop. 

Here are the execution times for both codes in seconds (vertical axis) for various value of \verb|n| (horizontal axis):

\begin{figure}[H]

\centering
\includegraphics[width=0.7\textwidth]{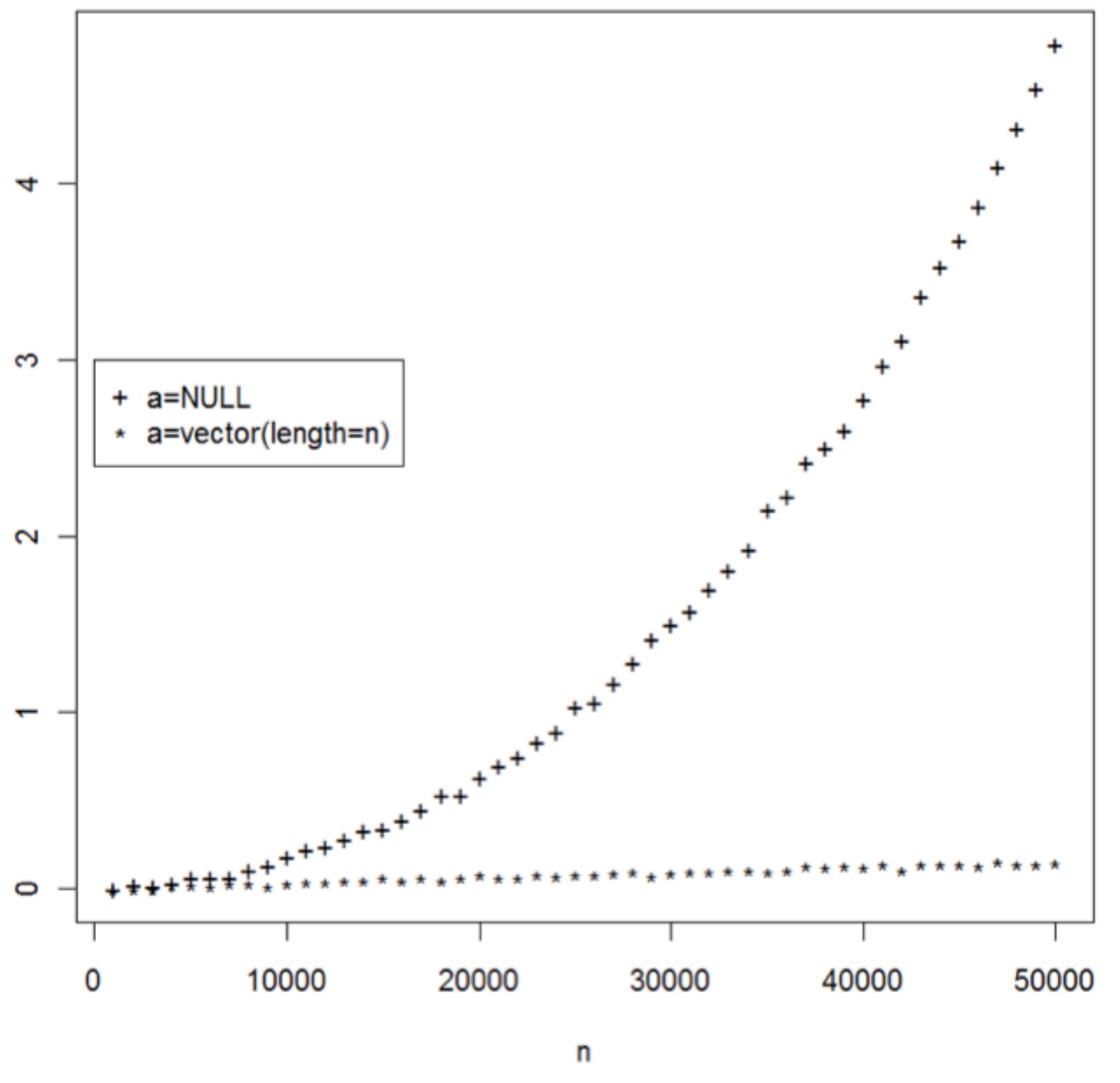} \\
\caption{\textbf{Always set any R object to its final size as soon as this final size is known.}}
\end{figure}

Another mistake R users can make is to perform the same calculation over and over again in a loop. Do not use
\begin{Verbatim}[frame=single, samepage=true]
a=vector(length=n)
for(i in 1:n) a[i]=2*pi*sin(i)
\end{Verbatim}

use instead

\begin{Verbatim}[frame=single, samepage=true]
a=vector(length=n)
for(i in 1:n) a[i]=sin(i)
a=2*pi*a
\end{Verbatim}

In the first code, \verb|2*pi| is calculated over and over again, something unnecessary. In general, loops should always contain as few calculations as possible. Here are the execution times for the two codes in seconds (vertical axis) for various values of \verb|n| (horizontal axis):

\begin{figure}[H]
\centering
\includegraphics[width=0.7\textwidth]{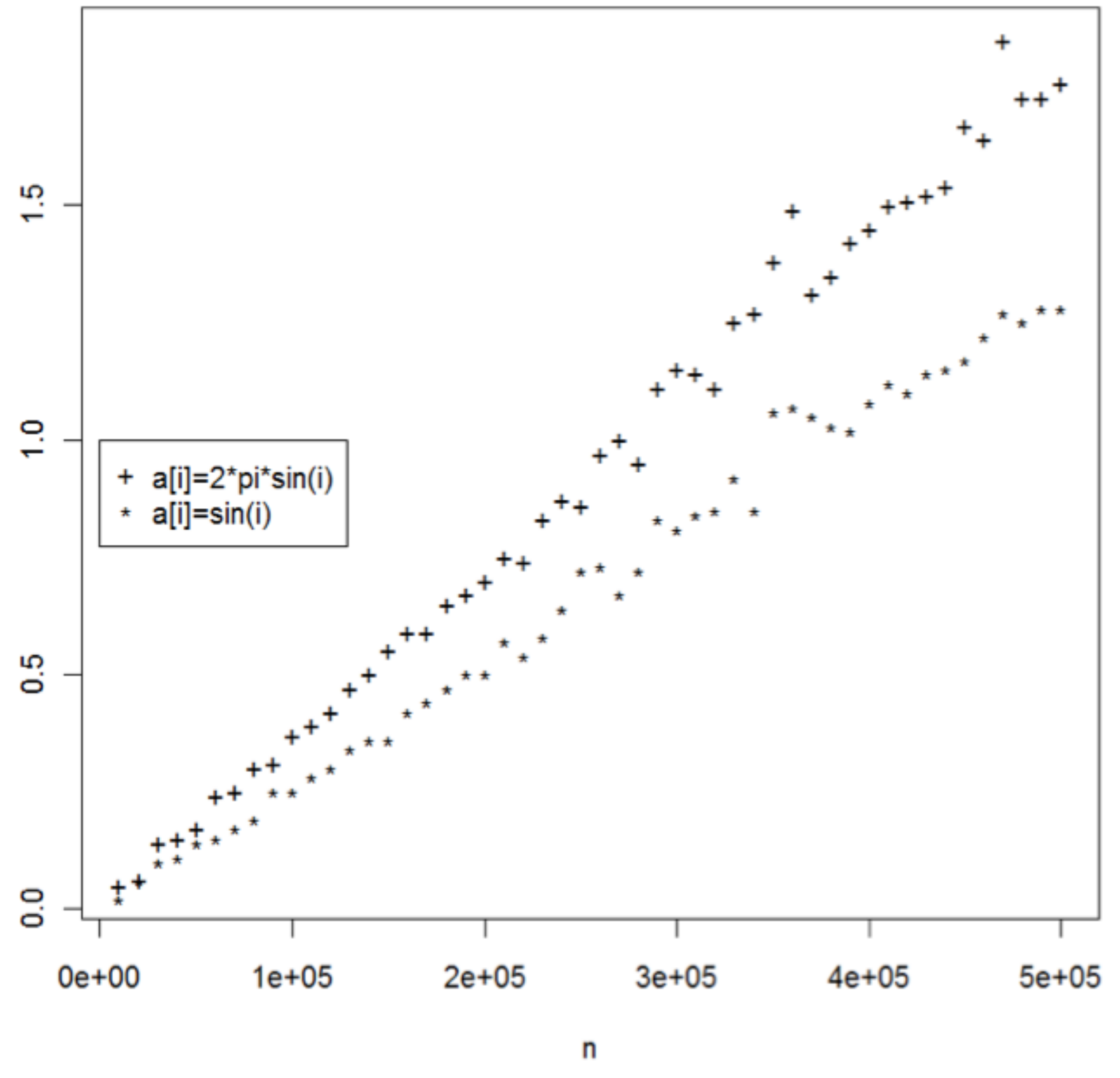} \\
\caption{Minimalist loops are better. But even better is to avoid loops!}
\end{figure}

Loops are actually something R does not like at all. Whenever you can avoid loops in R, just get rid of them. The difference between the execution time of a code including a loop and an alternative code without loop is usually impressive. Example:

\begin{Verbatim}[frame=single, samepage=true]
a=vector(length=n)
for(i in 1:n) a[i]=sin(i)
a=2*pi*a
\end{Verbatim}

versus

\begin{Verbatim}[frame=single, samepage=true]
a=2*pi*sin(1:n)
\end{Verbatim}

Both codes give the same result, \verb|a|. Here are the execution times in seconds for various values of \verb|n|:

\begin{figure}[H]
\centering
\includegraphics[width=0.55\textwidth]{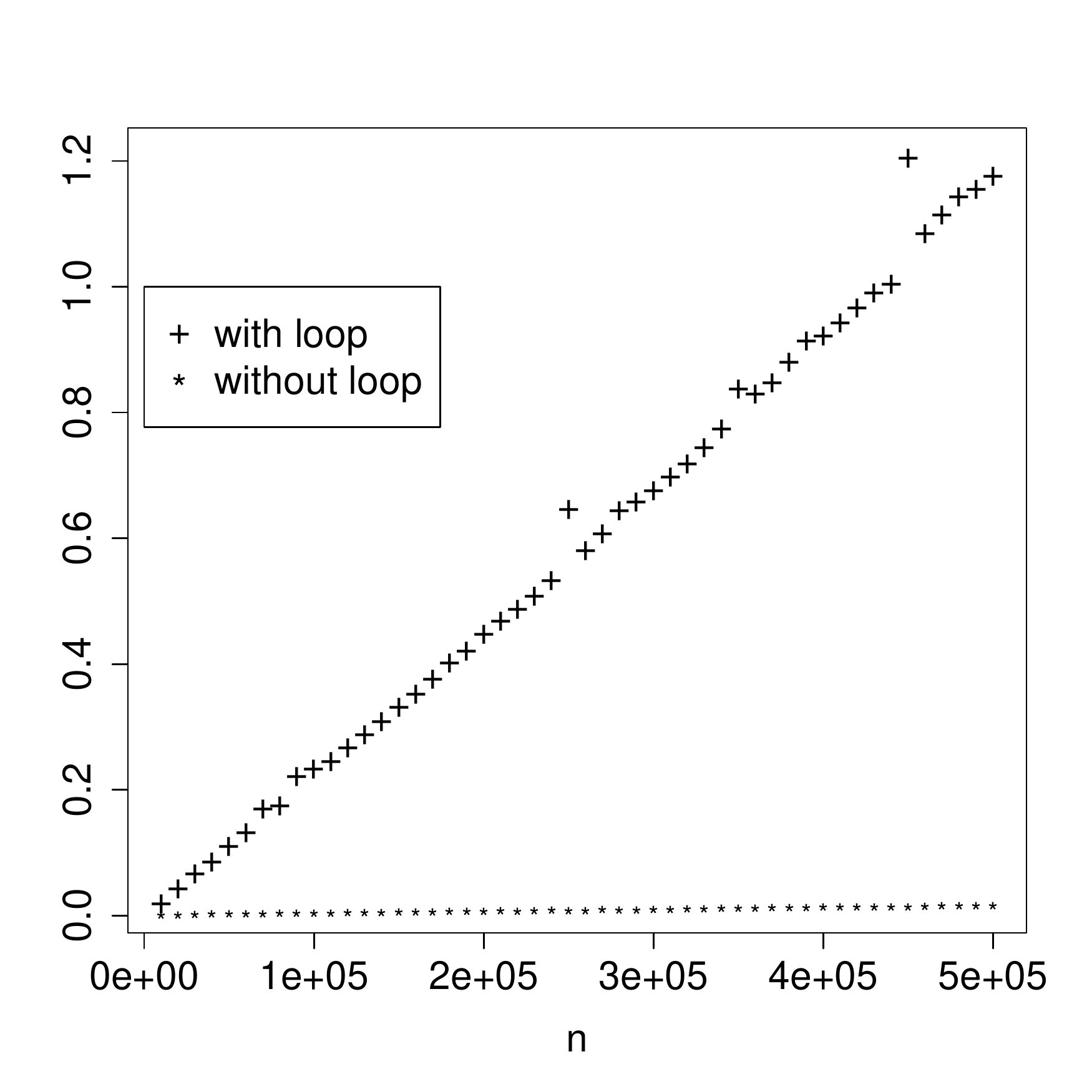} \\
\caption{R hates loops.}
\end{figure}

The second code is faster by a large amount. Moreover, it is also much easier to understand. Remember: R works better with vectors, matrices and tables. Loops are, in general, more expensive, as well as dataframes and lists.

There is a price to pay when you get rid of loops: the RAM is put under more pressure. In R, loops typically put a lot of stress on the CPU. By getting rid of them, the pressure is pushed toward the RAM. If you use a computer with a limited amount of RAM, you might be better off using more loops to avoid running out of RAM! Nowadays (2015) most computers have at least 4 GBs of RAM, which is more than most R codes will ever use. In short: get rid of loops whenever you can and do not worry too much about the effect on random-access memory.

Getting rid of loops is not always easy, especially for the unexperienced R user. For this reason, a practical example is shown hereafter.

Original code using a loop:

\begin{Verbatim}[frame=single, samepage=true]
n=10
y=rnorm(n)
W=vector(length=n)
for(i in 1:n) W[i]=sum(y[1:i]<y[i])
\end{Verbatim}

Code leading to the same result without loop:

\begin{Verbatim}[frame=single, samepage=true]
n=10
y=rnorm(n)
W=matrix(1:n, nrow=n, ncol=1)
W=apply(W, 1, function(x) sum(y[1:x]<y[x])) 
\end{Verbatim}

Notice how in this second code the \verb|i| index has been replaced by \verb|x| and how \verb|W| is used as a matrix of integers of size \verb|n|, the elements of which being referred to by \verb|x|. 

Last time both codes were run by the author of this paper, the execution times were 72.408 seconds versus 0.024 second.

\subsection{What part of my code is slow?}

When facing a large amount of slow code and knowing R hates loops, getting rid of all of them could be seen as the best thing to do. But getting rid of loops can be a time-consuming process. Most importantly, loops are not always the reason for a slow code!

Before any attempt to improve a slow code is made, the user should therefore first identify what part(s) of the code is(are) worth working on to avoid wasting valuable time rewriting code that was not a problem to start with. \cite{wickham2014advanced}:

\begin{quote}
It's easy to get caught up in trying to remove all bottlenecks. Don't! Your time is valuable and is better spent analysing your data, not eliminating possible inefficiencies in your code. Be pragmatic: don't spend hours of your time to save seconds of computer time. To enforce this advice, you should set a goal time for your code and optimise only up to that goal. This means you will not eliminate all bottlenecks.
\end{quote}

The best currently available tool to help you out in finding bottlenecks worth working on is the function \verb|Rprof| (\citealp{Rmanual}). Here is how to use it:

\begin{Verbatim}[frame=single, samepage=true]
Rprof("profiling.out")

### Your code ###

Rprof()
summaryRprof("profiling.out")
\end{Verbatim}

As an example, the following code will be used:

\begin{Verbatim}[frame=single, samepage=true]
data1=rnorm(10^7)

for(i in 1:(length(data)-1)) {
	data1[i]=data1[i]+data1[i+1]
	data1[i]=exp(data1[i]^2)
}

data2=rnorm(10^7)

data1=data2*data1

matrix.data1=matrix(data1, nrow=100, byrow=TRUE)
matrix.data2=matrix(data2, nrow=100, byrow=TRUE)

almost.final.result=matrix.data1%*%t(matrix.data2)

final.result=solve(almost.final.result)
\end{Verbatim}

The result of the profiling is

\begin{Verbatim}[frame=single, samepage=true]
> summaryRprof("profiling.out")
$by.self
            self.time self.pct total.time total.pct
"rnorm"          2.02    56.42       2.02     56.42
"%*%"            1.16    32.40       1.16     32.40
"matrix"         0.20     5.59       0.20      5.59
"t.default"      0.14     3.91       0.14      3.91
"*"              0.06     1.68       0.06      1.68

$by.total
            total.time total.pct self.time self.pct
"rnorm"           2.02     56.42      2.02    56.42
"%*%"             1.16     32.40      1.16    32.40
"matrix"          0.20      5.59      0.20     5.59
"t.default"       0.14      3.91      0.14     3.91
"t"               0.14      3.91      0.00     0.00
"*"               0.06      1.68      0.06     1.68

$sample.interval
[1] 0.02

$sampling.time
[1] 3.58

> |
\end{Verbatim}

\verb|$by.self| tells us that the function \verb|rnorm| eats up 56.42\% of the total execution time. Generating random numbers is what takes most of the time in the profiled code. \verb|%*%| requires 32.40\% of the total execution time. In other words: 32.40\% of the execution time is spent performing matrix multiplication. The loop in the code does not appear to be an issue.

You can define how precise the profiling of a code should be using the \verb|interval| argument:

\begin{Verbatim}[frame=single, samepage=true]
Rprof("profiling.out", interval=0.02)

### Your code ###

Rprof()
summaryRprof("profiling.out")
\end{Verbatim}

Default value is 0.02. The smaller the value, the better. However a small value relative to the total execution time of the code can make everything crash. The default value of 0.02 might not even work for a code requiring, say, one hour to run. If you encounter any issue, increasing the value to maybe 10 or even 100 should fix the problem.

For most codes, using \verb|Rprof| as just shown is enough to quickly identify bottlenecks. However, if the amount of code to profile is very large, using \verb|Rprof| as just shown might not be helpful. Indeed, if \verb|Rprof| is reporting that the function \verb|sum| eats up 90\% of the total execution time for a very large code where the function \verb|sum| is used every 5 lines, a critical question remains unanswered: what instances of the \verb|sum| function should be worked on? It is indeed very likely only a few instances of \verb|sum| are an issue in the code, not all of them.

For such cases, the execution time for \textit{each line of code} is needed. Such line profiling of your code is possible using \verb|Rprof| again ; you just need to set the \verb|line.profiling| argument TRUE and to wrap up the code to profile as a function:

\begin{Verbatim}[frame=single, samepage=true]
Rprof("profiling.out", line.profiling=TRUE)
 
### An R function ###

Rprof()
summaryRprof("profiling.out", lines="show")
\end{Verbatim}

As an example, let's wrap up the previously used code as an R function:

\begin{Verbatim}[frame=single, samepage=true]
#1  my_R_function=function() {
#2    data1=rnorm(10^7)
#3    for(i in 1:(length(data)-1)) {
#4      data1[i]=data1[i]+data1[i+1]
#5      data1[i]=exp(data1[i]^2)
#6    }
#7
#8    data2=rnorm(10^7)
#9    data1=data2*data1
#10
#11    matrix.data1=matrix(data1, nrow=100, byrow=TRUE)
#12    matrix.data2=matrix(data2, nrow=100, byrow=TRUE)
#13
#14    almost.final.result=matrix.data1%*%t(matrix.data2)
#15
#16    final.result=solve(almost.final.result)
#17  }
\end{Verbatim}

and let's line profile it:

\begin{Verbatim}[frame=single, samepage=true]
Rprof("profiling.out", line.profiling=TRUE)
 
my_R_function()

Rprof()
summaryRprof("profiling.out", lines="show")
\end{Verbatim}

in order to get as result

\begin{Verbatim}[frame=single, samepage=true]
$by.self
    self.time self.pct total.time total.pct
#2       1.10    30.90       1.10     30.90
#8       1.08    30.34       1.08     30.34
#14      1.04    29.21       1.04     29.21
#12      0.16     4.49       0.16      4.49
#11      0.14     3.93       0.14      3.93
#9       0.04     1.12       0.04      1.12

$by.line
    self.time self.pct total.time total.pct
#2       1.10    30.90       1.10     30.90
#8       1.08    30.34       1.08     30.34
#9       0.04     1.12       0.04      1.12
#11      0.14     3.93       0.14      3.93
#12      0.16     4.49       0.16      4.49
#14      1.04    29.21       1.04     29.21

$sample.interval
[1] 0.02
\end{Verbatim}

This shows that lines \#2 and line \#8 (see self.pct) are eating up the same amount of time to be executed. Line \#14, the one related to matrix multiplication, is eating up 29.21\% of the total execution time.

Codes with only one or two lines of code eating up most of the execution time are the most promising codes for optimization. Hereafter is the line profiling of a code the author of this paper has been working on for two years:

\begin{Verbatim}[frame=single, samepage=true]
$by.self
                               self.time self.pct 
estimate.NAC.structure.of#27     70.29    99.97     
estimate.NAC.structure.of#63      0.02     0.03 

$sample.interval
[1] 0.01

$sampling.time
[1] 70.31
\end{Verbatim}

Only one line of code turned out to account for 99.97\% of the total execution time. Rewriting that line of code in C (which is the topic of the next subsection) allowed to reduce the total execution time of the related function by more than 80\%.

\subsection{Writing part of your code in C}

Once a line of code eating up most of the execution time has been identified, rewriting it in a different way in order to go faster might actually make things worse. In such case, there is still something the user can try to speed up the code. The line of code (and if necessary, its surroundings) might go faster if rewritten as a C function.

C is usually much faster than R, especially with loops. If you are not used to C, there are excellent tutorials out there, simply search for them using Google.

How to proceed is shown hereafter through a first example, extracted from \cite{peng2002introduction}.

Given a sample of $n$ independent and identically distributed observations $x_1, \ldots, x_n$, the density $f(x)$ from which the sample was drawn can be estimated at an arbitrary point $x$ using

\begin{equation*}
\widehat{f(x)}=\frac{1}{nh}\sum^n_{i=1}K \left(\frac{x-x_i}{h}\right)
\end{equation*}
where $K$ is a kernel function, for example

\begin{equation*}  \label{build_back_H}
K(z)=\frac{1}{\sqrt{2\pi}}e^{-z/2}
\end{equation*}

and $h$ is the bandwidth.

An R implementation of this is, with \verb|xpts| being a vector of arbitrary points where we want to estimate the density and \verb|data| the vector containing the observations $x_1, \ldots, x_n$:

\begin{Verbatim}[frame=single, samepage=true]
ksmooth1 <- function(data, xpts, h) {
        dens <- double(length(xpts))
        n <- length(data)
        for(i in 1:length(xpts)) {
                ksum <- 0
                for(j in 1:length(data)) {
                        d <- xpts[i] - data[j]
                        ksum <- ksum + dnorm(d / h)
                }
                dens[i] <- ksum / (n * h)
        }
        return(dens)
}
\end{Verbatim}

The execution time for this function with an input sample of size \verb|n=500| from a $\chi^2$ distribution and a vector \verb|xpts| of size 10000 is:

\begin{Verbatim}[frame=single, samepage=true]
> data=rchisq(500, 3)
> xpts=seq(0, 10, length=10000)
> h=0.75
> system.time({
+ fx_estimated=ksmooth1(data, xpts, h)
+ })[[3]]
[1] 18.33
\end{Verbatim}

In order to improve this execution time, the \verb|ksmooth1| function is going to be rewritten in C. The result is

\begin{Verbatim}[frame=single, samepage=true]
#include <R.h>
#include <Rmath.h>
void kernel_smooth(double *data, int *n, double *xpts, 
               int *nxpts, double *h, double *result){
        int i, j;
        double d, ksum;
        for(i=0; i < *nxpts; i++) {
                ksum = 0;
                for(j=0; j < *n; j++) {
                        d = xpts[i] - data[j];
                        ksum += dnorm(d / *h, 0, 1, 0);
                }
        result[i] = ksum / ((*n) * (*h));
        }
}
\end{Verbatim}

The header \verb|#include <Rmath.h>| conveniently allows to use the R \verb|dnorm| function in C.

The above code must be stored in a .c file, for instance \verb|my_c_function.c|, then compiled into a .dll (Windows) or .so (Mac, Linux) file. This new file is then loaded in R using:

\begin{Verbatim}[frame=single, samepage=true]
dyn.load("my_c_function.dll")
\end{Verbatim}

or

\begin{Verbatim}[frame=single, samepage=true]
dyn.load("my_c_function.so")
\end{Verbatim}

As a final step, the \verb|ksmooth2| function must be written in R. The purpose of this function is to call \verb|kernel_smooth|, to handle to it the raw data and to output the results to the user:

\begin{Verbatim}[frame=single, samepage=true]
ksmooth2 <- function(data, xpts, h) {
        n <- length(data)
        nxpts <- length(xpts)
        dens <- .C("kernel_smooth", 
                as.double(data), 
                as.integer(n), 
                as.double(xpts), 
                as.integer(nxpts), 
                as.double(h), 
                result = double(length(xpts)))
        return(dens[[6]])
}
\end{Verbatim}

A word about the \verb|.C| function in \verb|ksmooth2|. The \verb|.C| function takes for input arguments first the name (\verb|kernel_smooth|) of the C function located in the

\verb|my_c_function.dll/so|

\noindent file and next the arguments required to run that C function. The function \verb|.C| outputs a list containing the input arguments of the C function, \underline{possibly modified by that C function}. The input argument \verb|result| in the above code will indeed not be the same before and after the \verb|kernel_smooth| function is called.

\begin{Verbatim}[frame=single, samepage=true]
> data=rchisq(500, 3)
> xpts=seq(0, 10, length=10000)
> h=0.75
> system.time({
+ result=ksmooth2(data, xpts, h)
+ })[[3]]
[1] 0.37
\end{Verbatim}

\verb|ksmooth2|, which makes use of the C function \verb|kernel_smooth|, requires only 0.37 second to run while \verb|ksmooth1| requires about 18 seconds. This is a decrease of roughly 97\%.

Still not convinced? Hereafter is an extra example related to the work of \cite{segers2014nonparametric}.

Given iid observations $(x_{11}, x_{12}) \ldots (x_{n1}, x_{n2})$ coming from a bivariate density for which the related copula (\citealp{nelsen1999introduction}) is an Archimedean copula, we would like to estimate the Kendall cumulative distribution function related to that Archimedean copula.

In order to estimate the Kendall CDF, the first step is to compute the pseudo observations $w_1, ..., w_n$ using (\citealp{genest2011inference}):

\begin{equation*}
w_j=\frac{1}{n+1}\sum^n_{k=1}I\{x_{k1}<x_{j1}, x_{k2}<x_{j2}\}
\end{equation*}

The Kendall CDF is then estimated at an arbitrary point $t$ by 

\begin{equation*}
F_n(t)=\frac{1}{n}\sum^n_{i=1}I\{w_{i}\leq t\}
\end{equation*}

This last step can be performed using the \verb|ecdf| function in R. 

The goal here is to get the pseudo observations as fast as possible. Hereafter \verb|input.data| is a matrix with $n$ lines and $2$ columns and the pseudo observations are returned as a vector \verb|W|:

\begin{Verbatim}[frame=single, samepage=true]
get.pseudo.obs1=function(input.data) {
			n=length(input.data[,1])
			W=vector(length=n)
			W=apply(input.data, 1, 
			function(x){
				sum(input.data[,1]<x[1] 
				& input.data[,2]<x[2])
					})
			W=W/(n+1)
			return(W)
}
\end{Verbatim}

Using the R package copula (\citealp{copulaPackage}) to generate 10000 bivariate observations from a Gumbel Archimedean copula, the execution time for \verb|get.pseudo.obs1| is 

\begin{Verbatim}[frame=single, samepage=true]
> library(copula)
> theta=copGumbel@tauInv(0.5)
> model=onacopula("Gumbel", C(theta, c(1, 2)))
> n=10000
> input.data=rnacopula(n, model)
> system.time({
+ W=get.pseudo.obs1(input.data)
+ })[[3]]
[1] 5.47
> |
\end{Verbatim}

Let's now calculate the pseudo observations in C, where \verb|x| is the first column of \verb|input.data| and \verb|y| is the second column:

\begin{Verbatim}[frame=single, samepage=true]
#include <R.h>
#include <Rmath.h>

void Kendall(double *x, double *y, int *n, double *result){
  int i, j;
  double s;
  /* outer loop: for each i from 1 to n, compute w[i] */
  for(i=0; i < *n; i++) {
    s = 0;
    /* inner loop: count the number of points (x[j], y[j])
     that are dominated by (x[i], y[i]) */
    for (j=0; j < *n; j++) {
      if((x[j] < x[i]) && (y[j] < y[i])) {
				s += 1;
      }
    }
    result[i] = s/(*n+1);
  }
}
\end{Verbatim}

The above code is stored in a .c file, say \verb|kendall_function.c|, transformed into a .so or .dll file and then loaded in R issuing:

\begin{Verbatim}[frame=single, samepage=true]
dyn.load("kendall_function.dll")
\end{Verbatim}

or

\begin{Verbatim}[frame=single, samepage=true]
dyn.load("kendall_function.so")
\end{Verbatim}

Next, the \verb|get.pseudo.obs2| function is written in R. The purpose of this function is to call the \verb|Kendall| C function, to handle to it the raw data and to output the results to the user:

\begin{Verbatim}[frame=single, samepage=true]
get.pseudo.obs2=function(input.data) {
			n=length(input.data[,1])
			W=vector(length=n)
			x=input.data[,1]
			y=input.data[,2]
			W=.C("Kendall", 
				as.double(x), 
				as.double(y), 
				as.integer(n), 
				result = double(n))[[4]]
			return(W)
}
\end{Verbatim}

Running the function \verb|get.pseudo.obs2| with some data gives us an execution time of 0.47 second:

\begin{Verbatim}[frame=single, samepage=true]
> theta=copGumbel@tauInv(0.5)
> model=onacopula("Gumbel", C(theta, c(1, 2)))
> n=10000
> input.data=rnacopula(n, model)
> system.time({
+ w=get.pseudo.obs2(input.data)
+ })[[3]]
[1] 0.47
> |
\end{Verbatim}

The total execution time went down from 5.47 seconds to 0.47 second. This is a decrease of 91.44\%.

Besides having to learn a new language, one of the biggest challenge when writing C code is usually the transformation of the .c file into a .dll or .so file. The user usually needs to install a few things, depending of what his operating system is, before being able to convert a .c file into a .dll or .so file. 

What to do for Windows, Mac OS and Linux Ubuntu is reviewed hereafter.

First, Windows. The user should install Perl (get the 64-bit version) and Rtools. Download links:

\url{http://www.activestate.com/activeperl/downloads}

\url{http://cran.r-project.org/bin/windows/Rtools/index.html}

Restart your computer once both are installed. Create now a file named Rpath.bat (or any other name), this file containing only one line of text (for convenience, the content of the file has been broken in 5 lines hereafter):

\begin{Verbatim}[frame=single, samepage=true]
PATH=
C:\Rtools\gcc-4.6.3\bin;
C:\Rtools\bin;
C:\Perl64\bin;
C:\Program Files\R\R-2.15.1\bin\x64;
\end{Verbatim}

The four above paths should be changed depending on where Rtools, Perl and R were installed on the computer, but also depending what version of gcc (4.3.6 at the time of writing) and R (2.15.1 at the time of writing) were installed on the computer.

The file \verb|Rpath.bat| can be stored anywhere, but for convenience it should be stored in the folder containing the .c file intended to be compiled as a .dll file. For even more convenience, both the .c file and the .bat file should be stored in the \verb|C:/Users/your_username| folder. 

Next, press the Start Button an type \verb|cmd|:

\begin{figure}[H]
\centering
\includegraphics[width=0.6\textwidth]{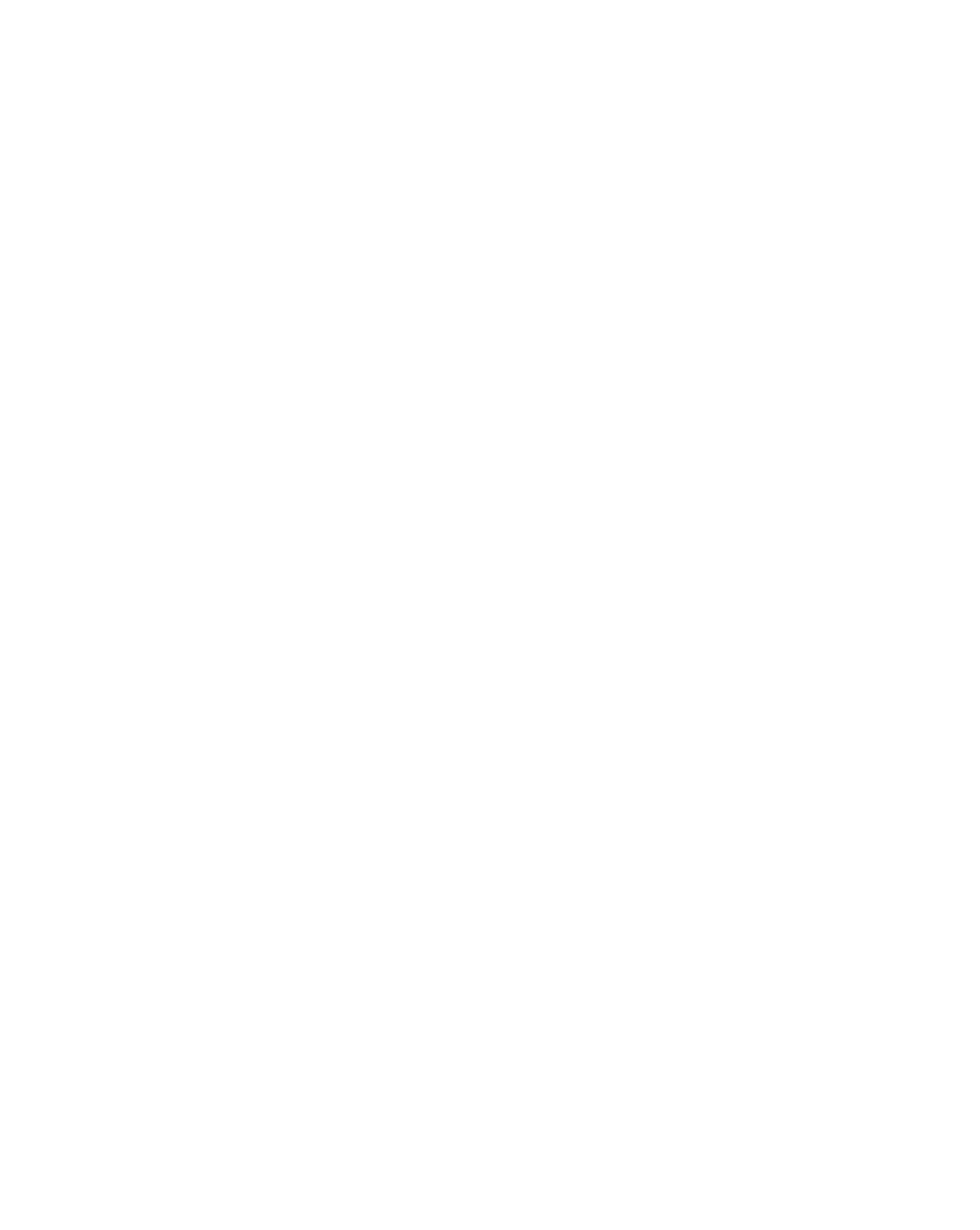} \\
\end{figure}

A black window will open.

\begin{figure}[H]
\centering
\includegraphics[width=0.55\textwidth]{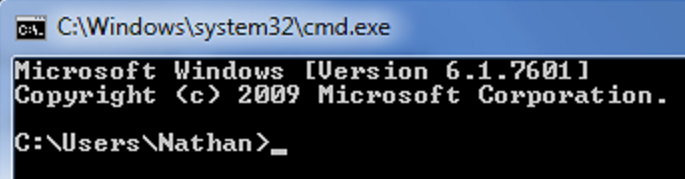} \\
\end{figure}

Submit the command \verb|Rpath.bat| and \verb|R CMD SHLIB kendall.c|, where \verb|kendall.c| is the name of your file to transform into a dll file. A .dll file should appears in the \verb|C:/Users/your_username| folder, ready for use in R.

\begin{figure}[H]
\centering
\includegraphics[width=0.55\textwidth]{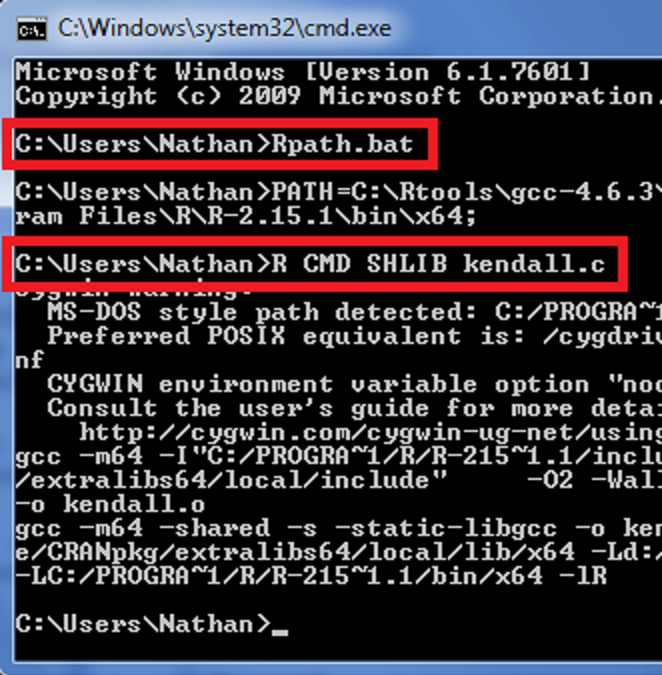} \\
\end{figure}

Second, Mac OS. The user needs to install something called the ``Command Line Tool for Xcode". First you need to register as an Apple Developer (this is just a formality). Link:

\url{https://developer.apple.com/register/}

It is very likely they will ask you to fill a far too long survey. Once you arrive at the developer page, browse it to find the appropriate version of Command Line Tool for your Mac. If you have no idea what version of Mac OS you are running, remember that all you have to do is click on the small apple icon top left of your screen and then pick ``About this Mac''.

\begin{figure}[H]
\centering
\includegraphics[width=0.3\textwidth]{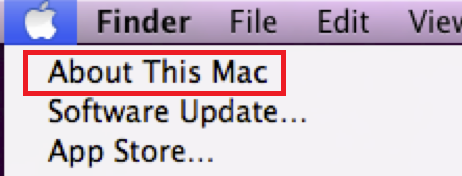} \\
\end{figure}

Once the .dmg file has been downloaded, run it to start the installation process. Restart your computer once the installation process is complete.

Put now your .c file into the \verb|/Users/your_username| folder for convenience. Open a terminal window. Use spotlight for this:

\begin{figure}[H]
\centering
\includegraphics[width=0.7\textwidth]{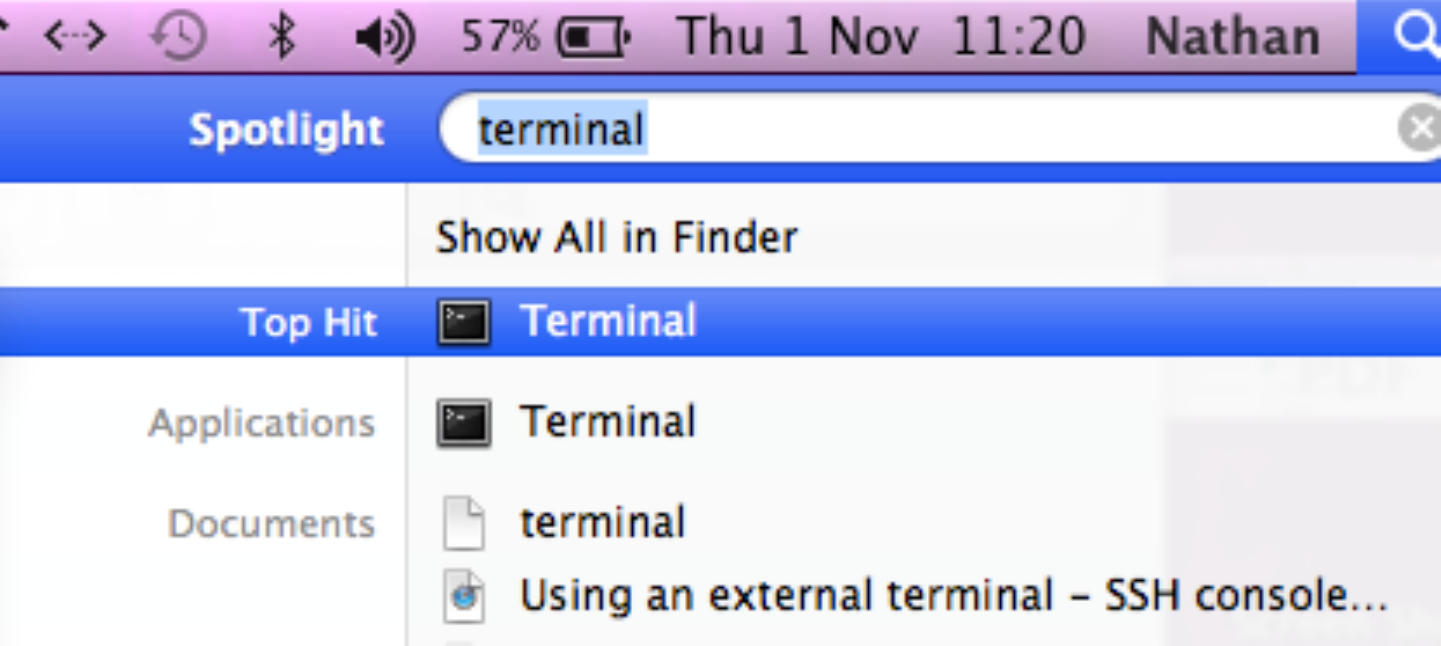} \\
\end{figure}

In the terminal, type \verb|R CMD SHLIB kendall.c|. A \verb|kendall.so| file should be created in the \verb|/Users/your_username| folder as a result.

\begin{figure}[H]
\centering
\includegraphics[width=0.85\textwidth]{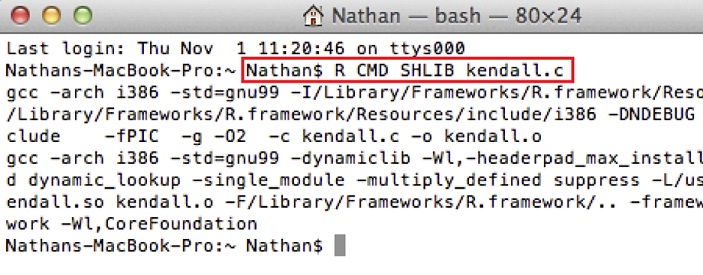} \\
\end{figure}

Last, Linux Ubuntu. Move your .c file into the \verb|/home/your_username| folder for convenience. Open a terminal window using the Dash:

\begin{figure}[H]
\centering
\includegraphics[width=0.35\textwidth]{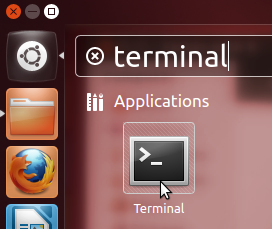} \\
\end{figure}

\begin{figure}[H]
\centering
\includegraphics[width=0.7\textwidth]{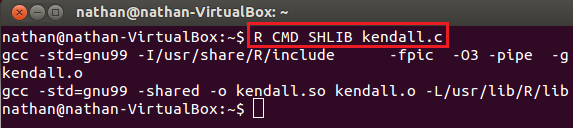} \\
\caption{\label{ubuntu_terminal_c}}
\end{figure}

Type \verb|R CMD SHLIB kendall.c| in the terminal window, as shown in Figure \ref{ubuntu_terminal_c}. A .so file should appear in the \verb|/home/your_username| folder. Linux is actually much more user-friendly for this than Mac OS or Windows.

\section{Using the others cores of your computer}

\subsection{Several R instances, one core for each \label{several_cores_one_instance}}

R is not natively able to use more than one core at the same time. Most computers today have at least two cores, usually 4 or 8.

A very effective way to make use of these extra cores is to use several R instances at the same time. The operating system will indeed always assign a different core to each new R instance. 

How to open several R instances on your computer?

On Windows 7, just click the R icon you have on your desktop or in your Start Menu several times. On Windows 8, clicking several times on the R icon you have on the desktop works, but not using the new Start Menu.

On Mac OS, clicking several times on the R icon will not work. Instead, you have to open several times the Terminal and then you have to open an R instance within each Terminal window. Start by opening the Terminal once, using Spotlight.

\begin{figure}[H]
\centering
\includegraphics[width=0.6\textwidth]{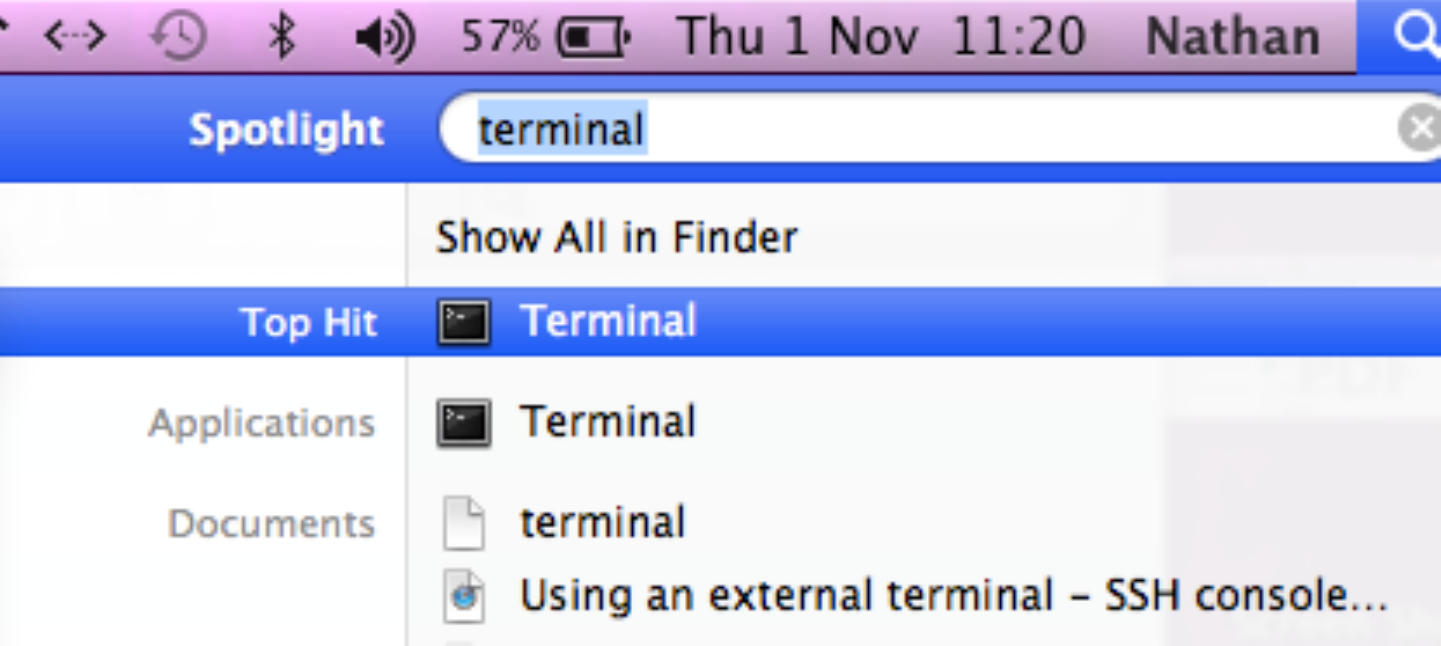} \\
\end{figure}

With one Terminal window opened, press \verb|cmd + N| to open a new one. Do this as many times as needed to get several Terminal windows. Next, open R within each Terminal window by typing the letter \verb|R| and pressing enter. If you want to open 64-bit R (assuming typing R in the terminal did not already lead to a 64-bit instance of R), try typing \verb|R64| instead.

\begin{figure}[H]
\centering
\includegraphics[width=0.9\textwidth]{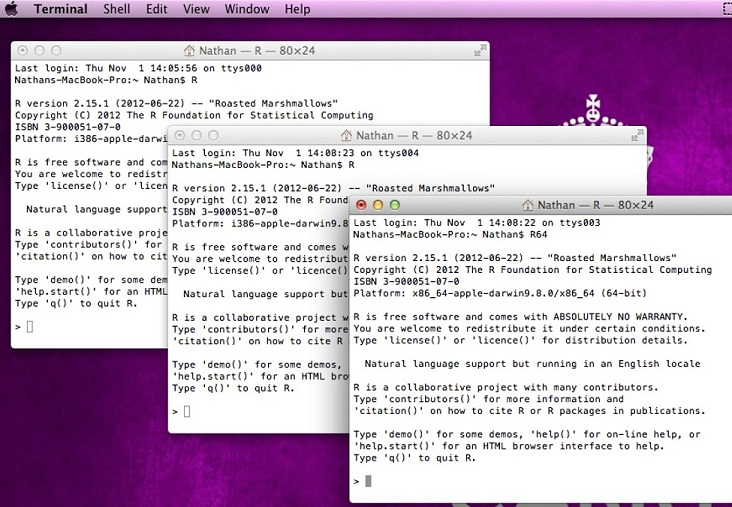} \\
\caption{3 R instances on a Mac}
\end{figure}

Last for Ubuntu Linux, just use the Dash several times to open several times the Terminal. Then within each Terminal, type \verb|R| to open R.

What happens if more R instances than the number of available cores are opened? Some instances will compete for the same core, slowing down calculations. The user should also keep in mind that many other processes are running on the computer and need access to the CPU, too. If a web browser is opened while the number of active R instances is exactly equal to the total number of cores available, at least one R instance will start to compete with the web browser for the same core, which can slow down both the browser and the calculations performed by the R instance.

If the total number of core available on the computer is $n_c$, a good rule of thumb is to always leave one core free and to open a maximum of $n_c-1$ R instances to avoid any interference between R processes and other processes on the computer.

With $n_c-1$ active R instances, open Windows Task Manager, System Monitor or Activity Monitor. Submit the following code to each R instance and observe what happens:

\begin{Verbatim}[frame=single, samepage=true]
for(i in 1:9000) {
  data=rnorm(10^5)
}
\end{Verbatim}

Each of the $n_c-1$ R instances should need about 100 seconds to be done. If you are using a laptop, expect the fans to start spinning at maximum speed and to make a lot of noise. Using $\frac{n_c-1}{n_c}\times100\%$ of the total computing power of the computer produces a lot of heat and also requires more energy: the level of the battery of the computer if it has one and is not in charge mode should start dropping fast.

A year ago, the author of this paper was asked an unexpected question: is it safe to use nearly 100\% of the total computing power for an extended period of time? The answer to that question is unfortunately beyond the scope of this paper and beyond the technical knowledge of the author of this paper. However, the author of this paper can report having repeatedly using nearly 100\% of the total computing power of many computers around him for extended period of times over the past few years without any problem: none of these computer died so far. Some common sense should however be applied to avoid trouble. Do not leave a computer running at 100\% for a week in a very small space or in a box. Do not leave a computer running at 100\% in the sun. Do not leave a computer running at 100\% near a device producing heat. Do not stack computers that are working at 100\% if it can be avoided.

Having several active R instances at the same time is great to parallelize loops. If each iteration of a loop is independent from previous and future iterations, the loop can be broken down across the various R instances before gathering all results in one R instance.

Example: one wishes to estimate the distribution of the sample mean $\bar{X}_n$ when $n=100000$ and the raw data are iid observations from a Poisson distribution with arbitrary mean $\lambda$.

Solution: generate 100000 iid observations from the Poisson distribution, get the sample mean for that sample and repeat this process until you get, say, 90000 sample means (the more, the better). With these 90000 sample means, you can draw a histogram, allowing you to estimate the distribution of the sample mean $\bar{X}_n$ when $n=100000$ and the CDF of $X$ is a Poisson distribution. Because of the Central Limit Theorem (CLT), we can expect the distribution of $\bar{X}_n$ to be a normal distribution with mean $\lambda$ and variance $\lambda/n$. Let's check using one R instance:

\begin{Verbatim}[frame=single, samepage=true]
xbar=vector(length=90000)
n=100000
system.time({for(i in 1:90000) {
	data=rpois(n ,lambda=1)
	xbar[i]=mean(data)
}
})[[3]]
hist(xbar, breaks=100, freq=FALSE)
\end{Verbatim}

Getting the graph took about 450 seconds with the computer used at the time of writing. Can this number be decreased by using the other cores? Yes it can, but all the results must be gathered in one R instance in order to draw the histogram. Assuming $n_c-1=3$, it is suggested to define 2 active R instances among the 3 as slaves and one as the master. Once they are done, the 2 slaves will write down their results on the hard disk drive. The master will load these results, add its own results and draw the graph.

Code for the slaves:

\begin{Verbatim}[frame=single, samepage=true]
slave.nb=1 ### or 2, this is important!
xbar=vector(length=30000)
n=100000
system.time({for(i in 1:30000) {
	data=rpois(n ,lambda=1)
	xbar[i]=mean(data)
}
write.table(xbar, 
          file=paste("xbar", slave.nb, ".txt", sep=""), 
          col.names=FALSE, row.names=FALSE)
})[[3]]
\end{Verbatim}

For the master:

\begin{Verbatim}[frame=single, samepage=true]
xbar=vector(length=90000)
n=100000
system.time({for(i in 1:30000) {
	data=rpois(n ,lambda=1)
	xbar[i]=mean(data)
}

for(i in 1:2) {
xbar[(i*30000+1):(i*30000+30000)]=read.table(
         paste("xbar", i, ".txt", collapse="", sep=""),
         header=FALSE)[,1]
}
})[[3]]
hist(xbar, breaks=100, freq=FALSE)
\end{Verbatim}

With the computer used at the time of writing, each slave took about 140 seconds to be done, as well as the master. The same histogram as before was drawn about 3 times faster.

Should the master try to load the results from the slaves before the slaves are done, an error message will be returned. By submitting the code for the master last, this problem is likely to be avoided but could still occur. 

To make sure the master is not trying to load data from the HDD that are not there yet, one could let the slaves write down their status on the hard disk drive as well: the master will first read the status for each slave and if (and only if) all slaves are done, will load all the results from the slaves. If all the slaves are not done, the master will wait before trying again.

The new code for the slaves is

\begin{Verbatim}[frame=single, samepage=true]
slave.nb=1 ### or 2, this is important!
status="running"
write.table(status, paste("status", slave.nb, ".txt", sep=""), 
col.names=FALSE, row.names=FALSE)
xbar=vector(length=30000)
n=100000
system.time({for(i in 1:30000) {
	data=rpois(n ,lambda=1)
	xbar[i]=mean(data)
}
write.table(xbar, 
          file=paste("xbar", slave.nb, ".txt", sep=""), 
          col.names=FALSE, row.names=FALSE)
status="terminated"
write.table(status, paste("status", slave.nb, ".txt", sep=""), 
		col.names=FALSE, row.names=FALSE)
})[[3]]
\end{Verbatim}

and for the master

\begin{Verbatim}[frame=single, samepage=true]
xbar=vector(length=90000)
n=100000
system.time({for(i in 1:30000) {
	data=rpois(n ,lambda=1)
	xbar[i]=mean(data)
}
all.finished=FALSE
while(!all.finished) {
	status1=as.character(read.table("status1.txt", 
	                         header=FALSE)[1,1])
	status2=as.character(read.table("status2.txt", 
	                         header=FALSE)[1,1])
	if(status1=="terminated" & status2=="terminated") 
		all.finished=TRUE
	else {
		cat("Waiting for at least one slave \n")
		Sys.sleep(10)
				
	}
}
for(i in 1:2) {
xbar[(i*30000+1):(i*30000+30000)]=read.table(
         paste("xbar", i, ".txt", collapse="", sep=""),
         header=FALSE)[,1]
}
})[[3]]
hist(xbar, breaks=100, freq=FALSE)
\end{Verbatim}

Should one of the slave not be done yet when the master tries to load the related results, the master will go to sleep for 10 seconds before trying again.

Concluding remarks. Using several R instances at the same time is very effective, but there are nonetheless some issues with this approach:

\begin{itemize}
\item the risk of human error becomes high when there are too many R instances to deal with. Moreover, having to control everything ``by hand" over and over again can be annoying.
\item the hard disk drive is used to exchange data between R instances. This can significantly slow down everything if the amount of data to write and read is large.
\end{itemize}

Advantages:
\begin{itemize}
\item easy to understand, even for a newcomer.
\item very efficient, at least as long as the amount of data to exchange using the hard disk drive is low. Using 2 slaves and a master, we get the same result 3 times faster.
\end{itemize}

\subsection{One R instance, several cores}

Rather than opening several R instances at the same time, one could seek a way to use several cores within the same R instance. Many packages allow to do this, and the easiest ones to use are reviewed in this subsection.

\subsubsection{pnmath}

First, pnmath.

This package was created by Luke Tierney\footnote{luke@stat.uiowa.edu}. The last update was made during the summer of 2012. The package is still an experimental package and cannot be found on CRAN. 

To install this package, you need to download the file \verb|pnmath_0.0-4.tar.gz| using the below link:

\url{http://www.stat.uiowa.edu/~luke/R/experimental/}

Once the \verb|pnmath_0.0-4.tar.gz| file has been downloaded, submit in the R instance the command
\begin{Verbatim}[frame=single, samepage=true]
install.packages("pnmath_0.0-4.tar.gz", repos=NULL)
\end{Verbatim}

This will only work for Linux users. As far as the author of this paper knows, Windows and Mac OS users are unable to use the package.

Once the package is installed, simply load it in your R instance using
\begin{Verbatim}[frame=single, samepage=true]
library(pnmath)
\end{Verbatim}
and that's it! Your R instance will automatically use more than one core for any new calculation involving one of the following functions\footnote{This list does not include all the functions improved by pnmath}:
\begin{figure}[H]
\centering
\includegraphics[width=0.9\textwidth]{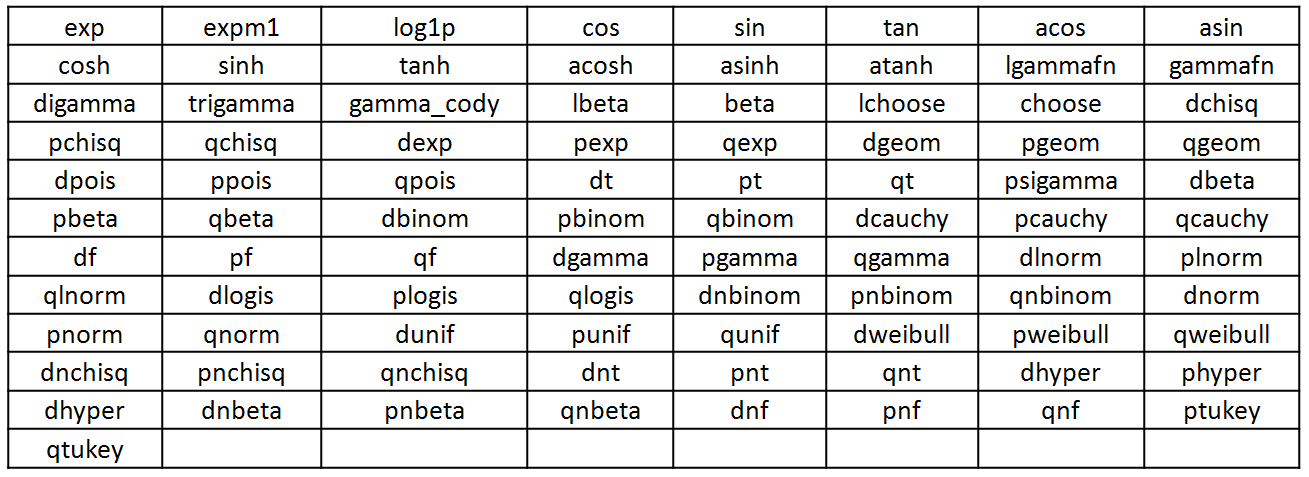} \\
\end{figure}
					
As an example, the following two lines of code have been run 50 times without and with the package \verb|pnmath| loaded:

\begin{Verbatim}[frame=single, samepage=true]
A=matrix(1:(10^7), nrow=1000)
B=tan(sin(cos(tan(A))))
\end{Verbatim}

\begin{figure}[H]
\centering
\includegraphics[width=0.65\textwidth]{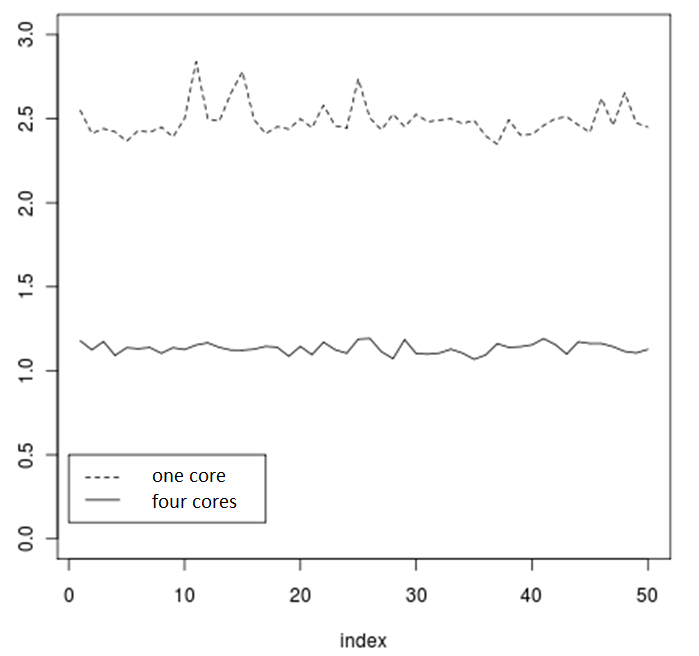} \\
\end{figure}

The vertical axis is the time required for each run in seconds. Without the package loaded, the active R instance only uses one core among the 4 available on the computer used for this experiment. When the package is loaded, up to 4 cores are available, although it is unclear whether or not \verb|pnmath| will use all of them. To better understand how \verb|pnmath| uses the total number of available cores, here is another graph where the package \verb|pnmath| was always loaded and the total number of available cores ranges from 1 to 8:

\begin{figure}[H]
\centering
\includegraphics[width=0.7\textwidth]{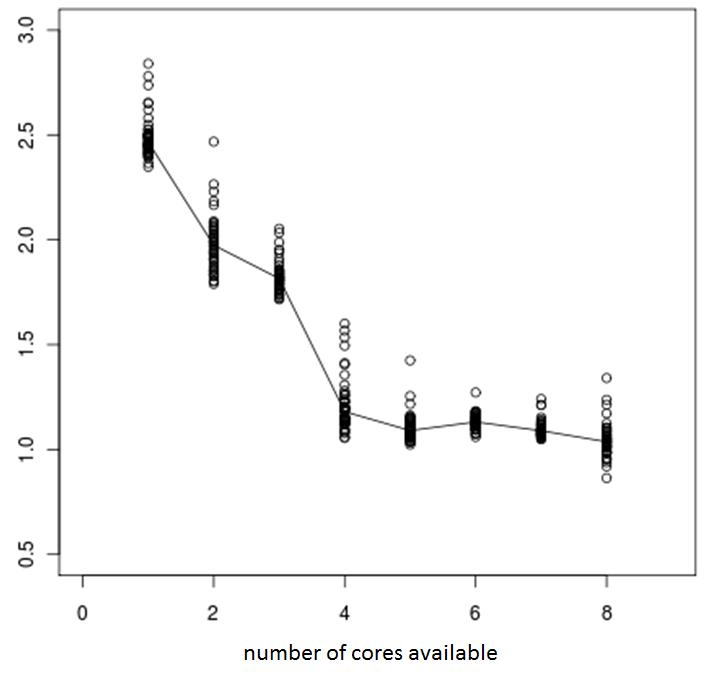} \\
\end{figure}

When there is only one core available, loading the package does not make any difference, as one would expect. Going from 2 to 3 cores does not help and it seems the package \verb|pnmath| fails to use more than 4 cores, since execution times do not improve anymore when the total number of available cores increases to become more than 4.

\subsubsection{pnmath0}

Second package of interest: \verb|pnmath0|. This is an alternative package to the \verb|pnmath| package, written by the same author. As it was the case for the \verb|pnmath| package, \verb|pnmath0| is not available on CRAN. To install the package, download the file \verb|pnmath0_0.0-4.tar.gz| using the link

\url{http://www.stat.uiowa.edu/~luke/R/experimental/}

and then submit in your R instance the command

\begin{Verbatim}[frame=single, samepage=true]
install.packages("pnmath0_0.0-4.tar.gz", repos=NULL)
\end{Verbatim}

Again, this won't work for Windows users or for Mac OS users.

To use this package, simply load it in the R instance. Many functions will go faster once the package has been loaded. The choice to use \verb|pnmath| or \verb|pnmath0| seems to be an arbitrary choice.

\subsubsection{doSMP}

Third, the packages doSMP and doMC. The doSMP package is intended to be used on Windows, the doMC package on Mac and Linux. There are some small differences between the two packages, they nonetheless have the same purpose: both packages allow to automatically break a \verb|for| loop into several parts that are then run on different cores at the same time. In other words: they allow to do what was done in subsection \ref{several_cores_one_instance} with significantly less trouble.

For instance, if a \verb|for| loop requires to run the same code 300 times, either doMC or doSMP will allow to run that code a 100 times on 3 different cores. This works only if successive iterations in the original loop are independent from one another.

Let's start with doSMP. Prior to the writing of this paper, the package was available on CRAN, all you had to do to get it was

\begin{Verbatim}[frame=single, samepage=true]
install.packages("doSMP")
\end{Verbatim}

Unfortunately, the package has been removed for CRAN. The developers of the package, Revolution Analytics, claimed\footnote{\url{http://heuristically.wordpress.com/2012/02/17/dosmp-removed-from-cran/}} it will be back for R 2.14.2 or maybe 2.15.0, but that was a while ago and R is way past 2.15.0 at the time of writing. It becomes therefore unclear whether or not this package will ever come back. Hereafter is shown how it worked prior to this removal from CRAN. If you intend to use doMC (this last package is still available on CRAN), knowing how things worked for doSMP can still be useful.

The \verb|for| loop to break down is:

\begin{Verbatim}[frame=single, samepage=true]
data=vector(length=1000)
for(i in 1:1000) {
	data[i]=sqrt(1/(sin(i))^2)-sum(rnorm(10^6))
} 
\end{Verbatim}

The doSMP package creates hidden slaves. Each slave will run part of the loop. To create 4 slaves, submit the following lines of code

\begin{Verbatim}[frame=single, samepage=true]
library(doSMP)
rmSessions(all.names=TRUE)
workers=startWorkers(4)
registerDoSMP(workers) 
\end{Verbatim}

These 4 slaves are only visible using the Windows Task Manager. Note: the R instance where you submit these lines is the master.

You are now ready to submit the \verb|for| loop to the master, but before you do so, you have to rewrite it a little:

\begin{Verbatim}[frame=single, samepage=true]
data=vector(length=1000)
data=foreach(i=1:1000) %dopar% {
	sqrt(1/(sin(i))^2)-sum(rnorm(10^6))
}
data=unlist(data)
\end{Verbatim}

Compare the rewritten loop with the original one. Both codes will output exactly the same result, a vector \verb|data| of size 1000. However in the rewritten loop, \verb|for| has become \verb|foreach|, \verb|%dopar%| appeared, etc.

Once the calculations are done, terminate the slaves issuing
\begin{Verbatim}[frame=single, samepage=true]
stopWorkers(workers)
\end{Verbatim}

Closing the master R instance will NOT terminate the slaves. If you closed the master R instance before closing the slaves, the only thing left to do is to use the Windows Task Manager to manually close the forgotten slaves, or to reboot the computer.

A final note: the package doSMP is more efficient when you use it to parallelize loops where each iteration requires a large execution time. As a result, if you have nested \verb|for| loops, always parallelize the most outer one.

\begin{figure}[H]
\centering
\includegraphics[width=0.5\textwidth]{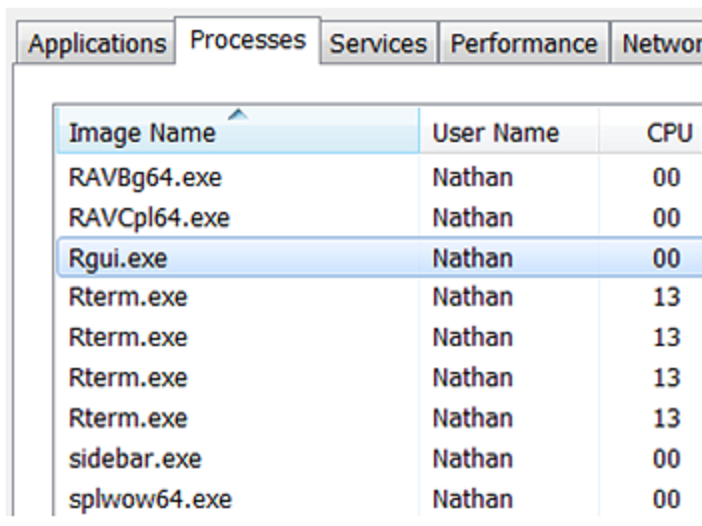} \\
\caption{Efficient parallelization. \label{efficient_loop}}
\end{figure}

You can check whether or not the target \verb|for| loop is efficiently parallelized by looking at the Windows Task Manager. A poor parallelization will result in slaves barely working and a master working hard, exactly as shown in Figure \ref{poor} below. An efficient parallelization will result in slaves working hard and a master doing nothing as shown in Figure \ref{efficient_loop}.

\begin{figure}[H]
\centering
\includegraphics[width=0.5\textwidth]{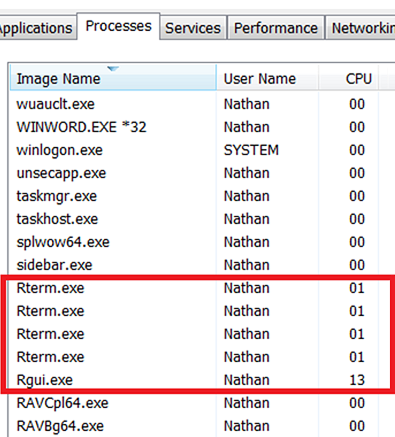} \\
\caption{4 slaves and one master. The parallelization is not efficient. \label{poor}}
\end{figure}

Try it yourself using the codes given in this subsubsection. Is the parallelization efficient? What do you suggest to improve the parallelization in case it is not efficient, keeping in mind the problem is usually that each individual iteration runs too fast?

\subsubsection{doMC}

This package (\citealp{doMC}) can be installed on both Mac OS and Linux by issuing the command

\begin{Verbatim}[frame=single, samepage=true]
install.packages("doMC")
\end{Verbatim}

This package works like the package doSMP, and most remarks made in the doSMP subsubsection are also true here. 

First you need to issue the following lines of code:

\begin{Verbatim}[frame=single, samepage=true]
library(doMC)
options(cores = 4)
registerDoMC()
\end{Verbatim}

This will not create 4 slaves as soon as the commands have been submitted, but only when some calculations are submitted.

Next, you need to rewrite your \verb|for| loop, exactly as it was done using doSMP:

\begin{Verbatim}[frame=single, samepage=true]
data=vector(length=1000)
data=foreach(i=1:1000) %dopar% {
	sqrt(1/(sin(i))^2)-sum(rnorm(10^6))
}
data=unlist(data)
\end{Verbatim}

doMC does not require you to worry about the slaves once the calculations are done. The slaves are automatically terminated once the calculations are done and created again if you submit new calculations. doMC does not allow slaves to exist if they do nothing, making this package easier to use than doSMP on Windows.

The following screenshot shows an example of parallelization using doMC on a Mac:

\begin{figure}[H]
\centering
\includegraphics[width=0.8\textwidth]{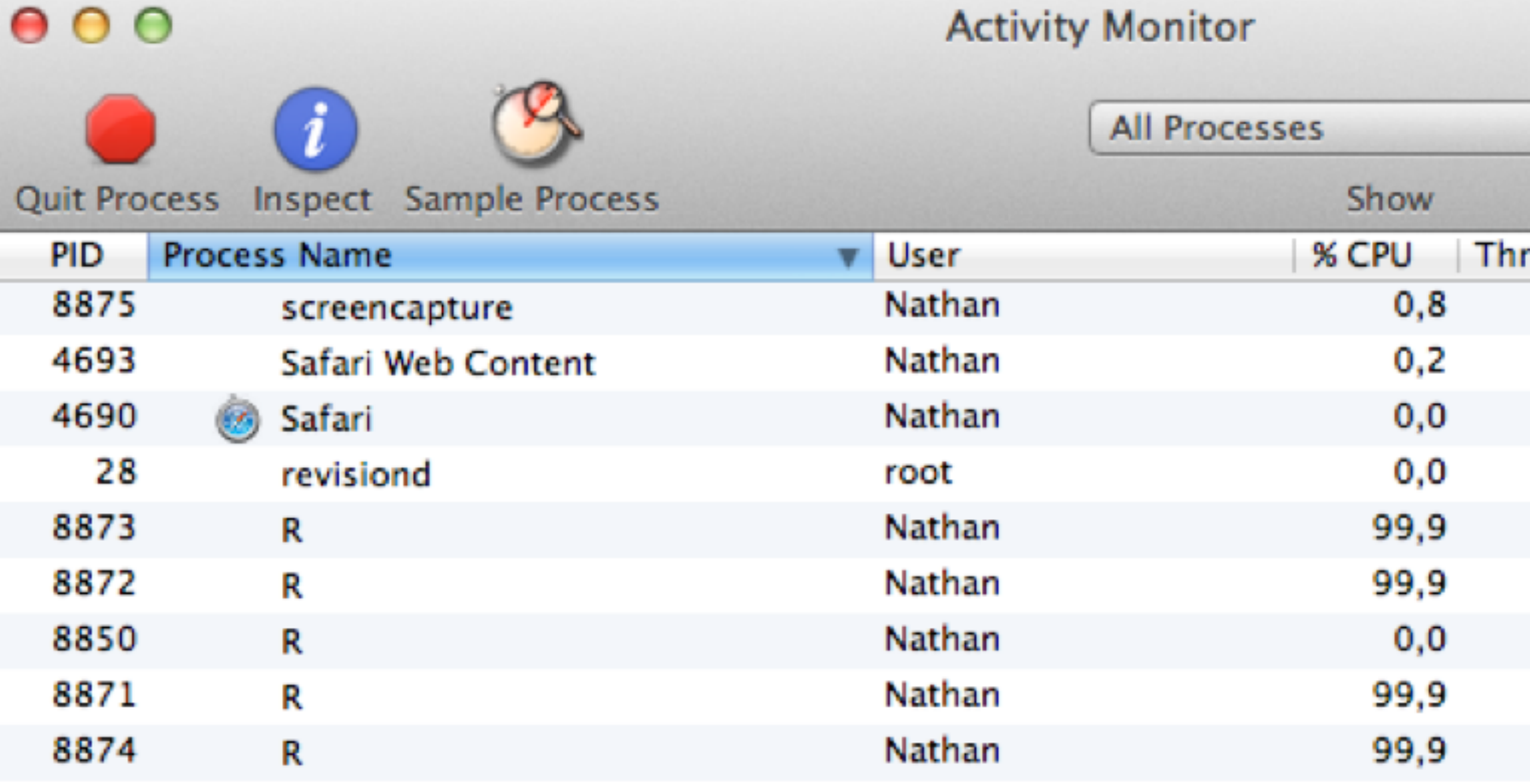} \\
\end{figure}

The parallelization is efficient, meaning each iteration of the parallelized \verb|for| loop is slow enough.

\subsubsection{parallel and snow}

Last, the snow and parallel packages. These packages can be used to run a function on several cores at the same time, for instance the \verb|runif| function. Start by loading the parallel package in your R instance with the command
\begin{Verbatim}[frame=single, samepage=true]
library(parallel)
\end{Verbatim}

This should work for Windows, Mac or Linux. No installation is required for this package.

Next load the package snow issuing

\begin{Verbatim}[frame=single, samepage=true]
library(snow)
\end{Verbatim}

Should the package not be installed, submit the command

\begin{Verbatim}[frame=single, samepage=true]
install.packages("snow")
\end{Verbatim}
before trying to load it again.

You are now ready to create slaves. For instance

\begin{Verbatim}[frame=single, samepage=true]
cl <- makeCluster(rep("localhost", 4), type = "SOCK")
\end{Verbatim}
will create 4 slaves. They can be seen using Windows Task Manager, Activity Monitor or System Monitor. They should not be working at the moment. 

Submit the command

\begin{Verbatim}[frame=single, samepage=true]
clusterCall(cl, runif, 10)
\end{Verbatim}
to ask each slave to generate 10 observations from the uniform distribution. 

\verb|clusterCall| always outputs the result as a list. The first element of the list is the output from the first slave, the second element is the output from the second slave, etc.

\begin{Verbatim}[frame=single, samepage=true]
> clusterCall(cl, runif, 10)
[[1]]
 [1] 0.52884968 0.90158064 0.96333475 0.04781719
 [5] 0.31112195 0.30859755 0.87067327 0.11359848
 [8] 0.53497501 0.30406015
[[2]]
 [1] 0.70646388 0.75440419 0.21298463 0.33821475
 [5] 0.22302076 0.62755126 0.69627268 0.04150676
 [8] 0.90047690 0.26068272
[[3]]
 [1] 0.84498905 0.43459654 0.75799121 0.62694915
 [5] 0.09589487 0.35845807 0.58648891 0.82221498
 [8] 0.48878884 0.19725313
[[4]]
 [1] 0.01429388 0.92020189 0.94640293 0.48377824
 [5] 0.44906074 0.39951102 0.19823086 0.27634168
 [8] 0.23072677 0.87001666
> |
\end{Verbatim}

Once you are done, shut down the slaves with the command

\begin{Verbatim}[frame=single, samepage=true]
stopCluster(cl)
\end{Verbatim}

Closing the R instance will also terminate the slaves.

\section{Using R on a cluster}

A cluster of computers consists of a set of loosely or tightly connected computers, usually in a same room kept under a specified temperature, usually stacked on one another.

Each computer of the set is called a node and usually has more cores than the average computer. Most of the time, each node is also equipped with a very large amount of RAM. Being able to run a slow R code on one or several nodes of a cluster is therefore worth the trouble: even if the user makes use of only one node of the cluster, he can expect to get his results faster than by using his own computer. Considering a cluster is made of many nodes, the total computing power available to the user is absolutely stunning. For instance, with a cluster, it becomes possible to break down a loop across hundreds of cores, using the same strategy explored in Subsection \ref{several_cores_one_instance}.

\begin{figure}[H]
\centering
\includegraphics[width=0.5\textwidth]{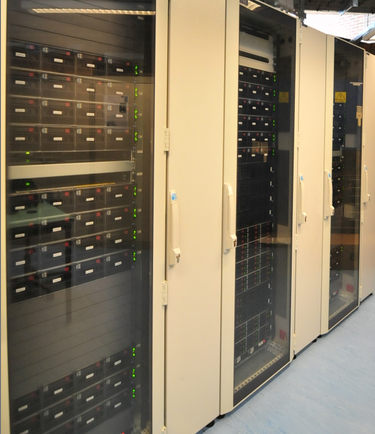} \\
\caption{a cluster located in Louvain-la-Neuve.}
\end{figure}

How exactly do you use a cluster? First, you need to connect to the frontend of the cluster. The frontend is nothing more than a node of the cluster, with a very special purpose: to manage the incoming users (you are of course not the only one using a given cluster). Once connected to the frontend, you will be able to jump toward another node and do whatever you want on this node (such as opening an R instance and submit some code in it). Most of the time however, the user is not allowed to freely jump from node to node across the cluster and will have to submit his code from the frontend to a job manager, such as Slurm (\citealp{yoo2003slurm}), which will decide where and when the code should be run on the cluster.

The purpose of this section is to show, step by step, how to run some R code and get back the results on a cluster through two examples.

Before going through these two examples, the reader needs to learn how to use a computer by mean of a Terminal window only (Mac, Linux) or by mean of Cygwin\footnote{Cygwin is a Terminal-like program for Windows} only (Windows). Here's how to install Cygwin:

Start by downloading setup.exe using the below link:

\url{http://www.cygwin.com/install.html}

Run setup.exe until you get to this:

\begin{figure}[H]
\centering
\includegraphics[width=0.7\textwidth]{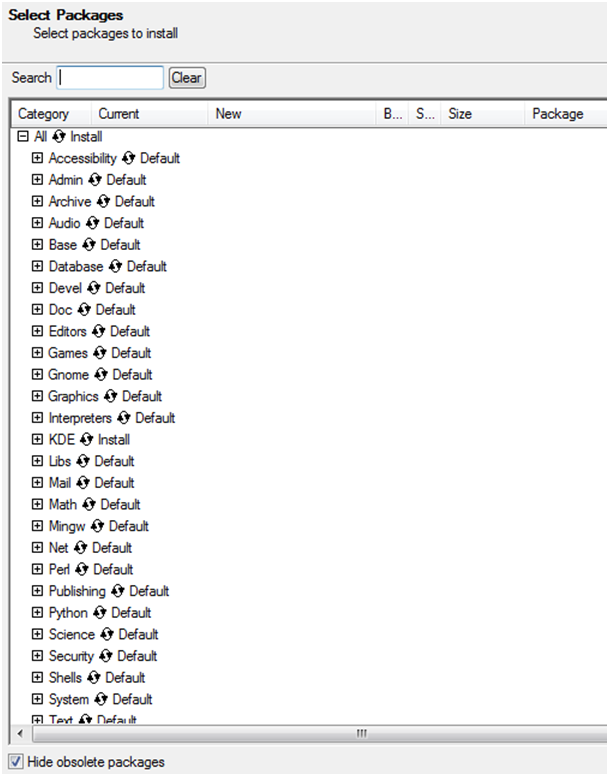} \\
\end{figure}

It is not required to install all available packages for Cygwin. Perform a search with the keyword \verb|ssh| and ask to install any package coming up. Do the same with the keyword \verb|scp|. This should be enough.

Restart the computer once the installation is complete. To run Cygwin, use the Start Menu:

\begin{figure}[H]
\centering
\includegraphics[width=0.4\textwidth]{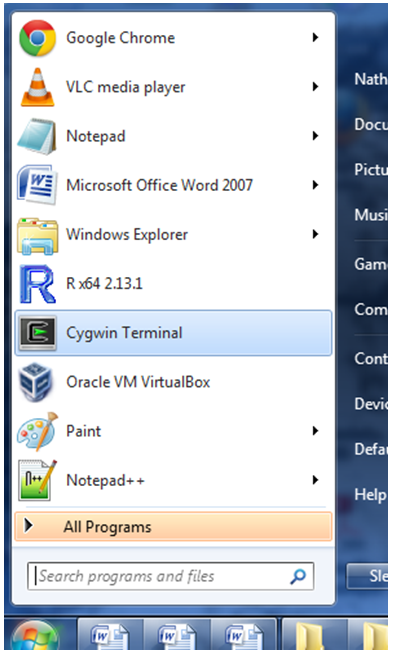} \\
\end{figure}

\subsection{Unix commands}

Open Cygwin or a Terminal window, depending of your OS. Linux users and Mac users already know at this point that they can run R from this window by simply typing the letter R (this will not work with Cygwin, though). Note: to quit R from a Terminal window and go back to the state the Terminal window was before typing the letter R, type in your R instance

\begin{Verbatim}[frame=single, samepage=true]
quit("no")
\end{Verbatim}

You can do much more than running R from a Terminal window. For instance, type

\begin{Verbatim}[frame=single, samepage=true]
pwd
\end{Verbatim}
to get the folder in which you are currently working (works whether you are using Cygwin or Terminal).

\begin{figure}[H]
\centering
\includegraphics[width=0.4\textwidth]{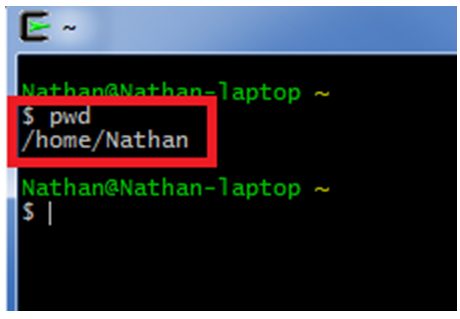} \\
\end{figure}

If you type
\begin{Verbatim}[frame=single, samepage=true]
ls -a
\end{Verbatim}

A list of everything in the working folder is returned:

\begin{figure}[H]
\centering
\includegraphics[width=0.8\textwidth]{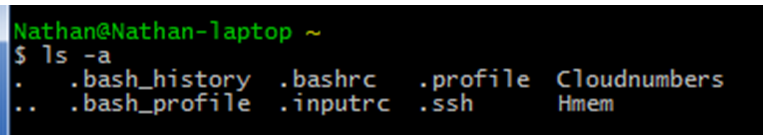} \\
\end{figure}

Note for Windows users: the folder \verb|/home/your_username| corresponds to \verb|C:\cygwin\home\your_username|, as you can see on the below picture:

\begin{figure}[H]
\centering
\includegraphics[width=0.9\textwidth]{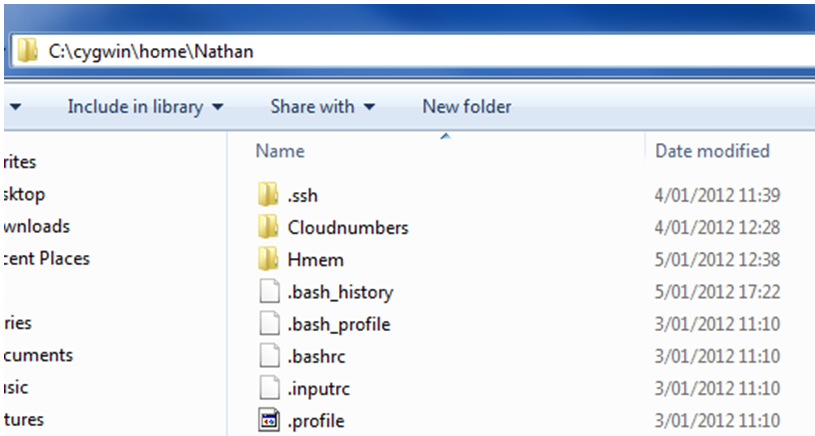} \\
\end{figure}

To change the working folder, use the command \verb|cd|. For instance, type

\begin{Verbatim}[frame=single, samepage=true]
cd ..
\end{Verbatim}

To move one level up from your current position
\begin{figure}[H]
\centering
\includegraphics[width=0.45\textwidth]{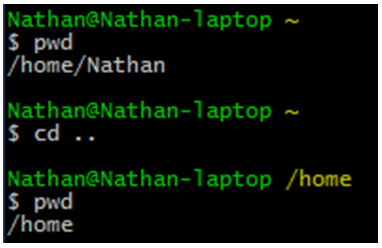} \\
\end{figure}

Type
\begin{Verbatim}[frame=single, samepage=true]
cd your_username
\end{Verbatim}

to go back to the \verb|your_username| folder.

To copy a file from one folder to the other, use

\begin{Verbatim}[frame=single, samepage=true]
cp /home/Nathan/script.r /home/
\end{Verbatim}

This command will copy the file \verb|script.r| from /home/Nathan/ to /home/

Note: if the working folder is already /home/Nathan/, the following commands will lead to the same result:

\begin{Verbatim}[frame=single, samepage=true]
cp /home/Nathan/script.r /home/
cp script.r /home/
cp ./script.r /home/
\end{Verbatim}

\verb|./| is indeed always replaced by the working folder.

To move a file from one folder to another, use the \verb|mv| command

\begin{Verbatim}[frame=single, samepage=true]
mv ./script.r /home/
\end{Verbatim}

To delete a file, use the \verb|rm| command (remove):

\begin{Verbatim}[frame=single, samepage=true]
rm /home/Nathan/script.r
\end{Verbatim}

To delete a folder, use

\begin{Verbatim}[frame=single, samepage=true]
rmdir /home/Nathan/
\end{Verbatim}

Note: you can only delete empty folders.

To create a folder, use

\begin{Verbatim}[frame=single, samepage=true]
mkdir Test
\end{Verbatim}

A folder named Test will be created in the working folder.

To see the content of a file, such as \verb|script.r|, type

\begin{Verbatim}[frame=single, samepage=true]
cat ./script.r
\end{Verbatim}

These commands are called Unix commands and there are many more. Most people today are not even aware they can interact with their computer using Unix commands. It is however important to master Unix commands when you use a cluster. Once the user is connected to a node of a cluster, Unix commands are usually the only way to interact with that node.

\subsection{First example: Hmem}

Hmem\footnote{http://www.ceci-hpc.be/clusters.html\#hmem} is a Belgian cluster managed by the ``Institut de calcul intensif et de stockage de masse", located at Louvain-la-Neuve, Belgium. This cluster is intended to be used by researchers from the French-speaking part of Belgium. As of 2015, the cluster comprises 17 nodes, a total of 816 cores and 512, 256 or 128 GBs of RAM per node.

To register (assuming you are a researcher working in the French-speaking part of Belgium), use the following link:

\url{http://www.ceci-hpc.be}

Once registration is completed, a small file (a key) will be sent to you by email. Copy this key in any folder you want, for instance 

\verb|/home/your_username/Hmem| 

The name of the file containing the key is \verb|id_rsa.ceci|:

\begin{figure}[H]
\centering
\includegraphics[width=0.8\textwidth]{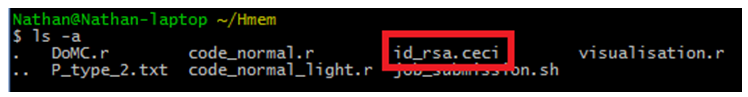} \\
\end{figure}

In order for this key to be usable, type

\begin{Verbatim}[frame=single, samepage=true]
chmod 600 /home/Nathan/Hmem/id_rsa.ceci
\end{Verbatim}

You can now connect to Hmem with the following command:
{\small
\begin{Verbatim}[frame=single, samepage=true]
ssh -i /home/Nathan/Hmem/id_rsa.ceci nuyttend@hmem.cism.ucl.ac.be
\end{Verbatim}
}

This will not work if your computer is not connected to the network of one of the French-speaking universities of Belgium.

\verb|nuyttend| was the username the author of this paper provided during his own registration. After the above command is submitted, a password is required. This password is the same that was provided during registration. Type now the command \verb|pwd| to know what is you working folder on the frontend:

\begin{figure}[H]
\centering
\includegraphics[width=0.9\textwidth]{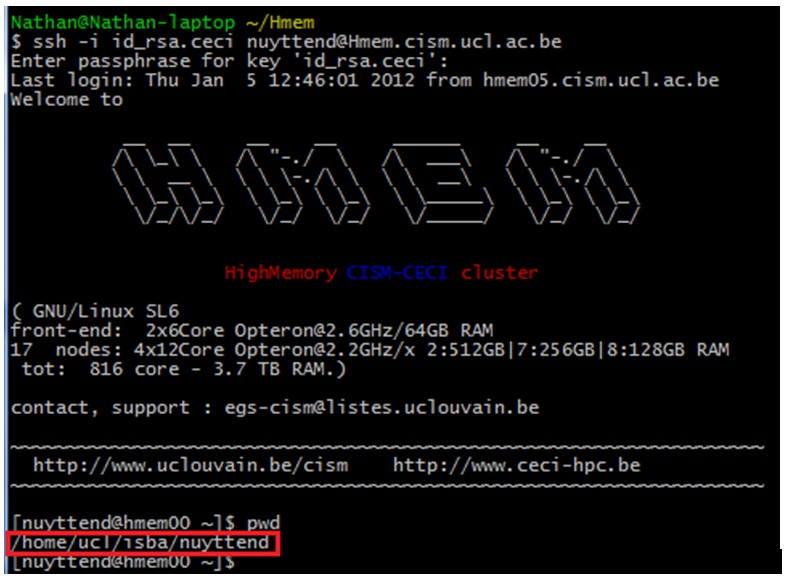} \\
\end{figure}

The working folder on the frontend is \verb|/home/ucl/isba/nuyttend|. If you want to jump from the frontend to another node, for instance node number 4, type

\begin{Verbatim}[frame=single, samepage=true]
ssh hmem04
\end{Verbatim}

Type now

\begin{Verbatim}[frame=single, samepage=true]
pwd
\end{Verbatim}

The working folder is still \verb|/home/ucl/isba/nuyttend/|. The node where the user is located does not matter: the working folder remains the same common folder everywhere, making things very easy.

To disconnect from the cluster, type exit. This sends you back on your own computer (that is, if you type the command \verb|pwd| again, the working folder will be a folder on your side).

How to send a file containing R code from your computer toward the cluster? Without being connected to the cluster, submit the following command:

{\scriptsize
\begin{Verbatim}[frame=single, samepage=true]
scp -i id_rsa.ceci script.r nuyttend@hmem.cism.ucl.ac.be:/home/ucl/isba/nuyttend
\end{Verbatim}
}

This will copy the file \verb|script.r| from your computer to the folder 

\verb|/home/ucl/isba/nuyttend| 

which is available from every node of the cluster. Both \verb|id_rsa.ceci| and \verb|script.r| are assumed to be in the same folder on the user's side before the above command is submitted.

Next, you want to run \verb|script.r| on the cluster. Beware special character in \verb|script.r| are usually not allowed, even as comments. To run \verb|script.r|, you need to speak with the job manager (called Slurm on Hmem) and tell it you want to run \verb|script.r| on one node of the cluster, how many cores you want, how much RAM and for how much time. If you want \verb|script.r| to be run on several nodes at the same time, simply ask several times the same thing to the job manager.

To speak with the job manager, you need to write a special file first. The name of this file is irrelevant. For this example, it will be called \verb|job_submission.sh|. Here is the content of the file:

\begin{Verbatim}[frame=single, samepage=true]
#!/bin/bash
#SBATCH --job-name=%j
#SBATCH --mail-user=me@gmail.com
#SBATCH --mail-type=ALL
#SBATCH --output=%j.tx
#SBATCH --time=02:00:00
# Acceptable time formats include "minutes", 
# "minutes:seconds", 
# "hours:minutes:seconds", "days-hours", "days-hours:minutes" 
# and "days-hours:minutes:seconds"
#SBATCH --ntasks=1
#SBATCH --nodes=1
#SBATCH --cpus-per-task=1
#SBATCH --mem-per-cpu=1024 
R CMD BATCH /home/ucl/isba/nuyttend/script.r
# end of job
\end{Verbatim}

\verb|me@gmail.com| should be replaced by your mail. This will allow Slurm to send you a mail when your script start to run but also when the calculations are terminated.

\verb|02:00:00| is the time you believe your script will need to be executed. With \verb|02:00:00|, you will get two hours. Note that whether or not your script is done after 2 hours is irrelevant: the R instance will be terminated, no matter what.

\verb|cpus-per-task=1| is the number of cores you need to use for your script. Unless you use some of the special R packages discussed above in this paper, you have no reason to ask for more than 1 core for your script.

\verb|mem-per-cpu=1024| is the memory you believe your script will require. You should always ask for a little more than what your script will really need, in order to avoid any problem.

\verb|R CMD BATCH /home/ucl/isba/nuyttend/script.r| should be changed according to where your script is located on the cluster.

Once the \verb|job_submission.sh| file is created, send it to the cluster using

{\tiny
\begin{Verbatim}[frame=single, samepage=true]
scp -i id_rsa.ceci job_submission.sh nuyttend@hmem.cism.ucl.ac.be:/home/ucl/isba/nuyttend
\end{Verbatim}
}

Connect now to the frontend and ask Slurm to read the .sh file with the command

\begin{Verbatim}[frame=single, samepage=true]
sbatch job_submission.sh
\end{Verbatim}
\verb|script.r| will then be queued by Slurm. You can check the queue with the command
\begin{Verbatim}[frame=single, samepage=true]
squeue
\end{Verbatim}

Once the script start to run, an email containing (among other things) the node number is sent. Feel free to jump toward that node and type

\begin{Verbatim}[frame=single, samepage=true]
top
\end{Verbatim}

To get a list of all processes currently running on that node. You should see at least one active R instance: yours.

Wait, where will you get the results of your script? Well, the only way to get the results of your script is to write them down on the hard disk drive of the cluster before the active R instance is terminated:

\begin{Verbatim}[frame=single, samepage=true]
setwd("/home/ucl/isba/nuyttend/")
*** main body of your R script ***
write.table(results, file="results.txt")
\end{Verbatim}

This way, the vector \verb|results| will be saved on the hard disk drive of the cluster in the folder \verb|/home/ucl/isba/nuyttend/|. To download the file \verb|results.txt| from the cluster, disconnect yourself from the cluster and type

{\tiny
\begin{Verbatim}[frame=single, samepage=true]
scp -i id_rsa.ceci nuyttend@hmem.cism.ucl.ac.be:/home/ucl/isba/nuyttend/results.txt /home/Nathan/Hmem
\end{Verbatim}
}

If you want to run the same script on several nodes at the same time, just submit the command

\begin{Verbatim}[frame=single, samepage=true]
sbatch job_submission.sh
\end{Verbatim}
several times. Keep in mind that if you run exactly the same version of \verb|script.r| on different nodes with

\begin{Verbatim}[frame=single, samepage=true]
setwd("/home/ucl/isba/nuyttend/")
*** main body of your R script ***
write.table(results, file="results.txt")
\end{Verbatim}
each node will replace the content of \verb|results.txt| by its own results. One way around this problem is to load the content of \verb|results.txt| first if the file exists, add to its content the results of the node, and then write everything down.

\subsection{Green}

Green was a Belgian cluster managed by the ``Institut de calcul intensif et de stockage de masse". It included 102 nodes for a total of 816 cores. This cluster was intended to be used by researchers from the ``Universit\'e Catholique de Louvain'' (UCL) only. Even if this cluster does not exist anymore, details on how to use this cluster remain useful and are given hereafter.

First, the user needs to register. Assuming you are a researcher from UCL, send a mail to someone from the ``Institut de calcul intensif et de stockage de masse"\footnote{http://www.uclouvain.be/73262.html}.

Once registered, you will not receive a key like it was the case with Hmem. Issuing the following command is enough to get access to the frontend of Green:

\begin{Verbatim}[frame=single, samepage=true]
ssh your_login@green.cism.ucl.ac.be
\end{Verbatim}

This will work provided your computer is connected to the network of UCL. 

Once the above command has been submitted, a weird message could ask you to type yes or no. Type yes and forget about it. Type now \verb|pwd| to find out what is your working folder on Green:

\begin{figure}[H]
\centering
\includegraphics[width=0.9\textwidth]{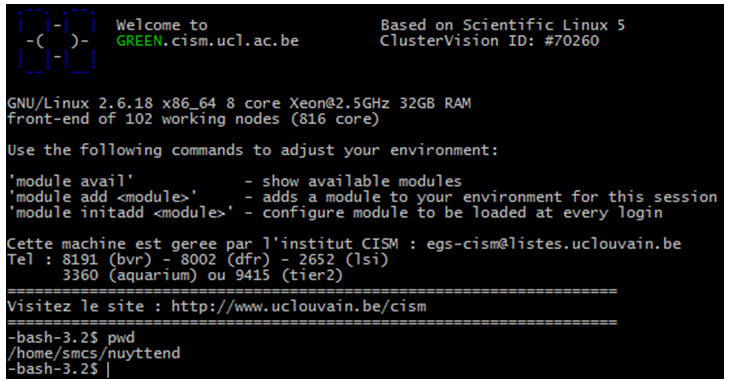} \\
\end{figure}

As you can see in the above screenshot, the working folder for the author of this paper was \verb|/home/smcs/nuyttend/|. Disconnect from the frontend issuing exit and send your R script to the cluster with the command

{\small
\begin{Verbatim}[frame=single, samepage=true]
scp script.r your_login@green.cism.ucl.ac.be:/home/smcs/nuyttend/
\end{Verbatim}
}

As it was the case for Hmem, you need to create a \verb|job_submission.sh| file and ask the job manager to read it. The job manager of Green is not Slurm but is called SGE. Here is the content of the \verb|job_submission.sh| file intended to be read by SGE:

\begin{Verbatim}[frame=single, samepage=true]
#!/bin/sh
#$ -N Name_of_the_job
#$ -pe mpich 1
# Advised: requested memory for each core
# 1G, 2G, 256M, etc
#$ -l h_vmem=512M
#$ -l mem_free=512M
#
#$ -l h_rt=60
# (xxxx sec or hh:mm:ss (max 5 days=120:0:0)
# SGE will kill your job after the requested period.
#
# Advised: your Email here, for job notification
#$ -M me@gmail.com
#$ -m bes
#
# Optional: ask for specific resources (licence, etc.) with 
## -l resourcename = ...
#
#$ -l nb=false
#
# Optional: activate resources reservation 
# when you need a large number of cores
## -R y
#
# Advised: output in the current working dir
#$ -cwd    
# Advised: combine output/error messages into one file
#$ -j y
#
# Launch job
echo "Got $NSLOTS slots. Temp dir is $TMPDIR, Node file is:"
cat $TMPDIR/machines
echo Start at
date
R CMD BATCH /home/smcs/nuyttend/script.r 
echo End at
date
# end of job
\end{Verbatim}

\verb|Name_of_the_job| is an arbitrary name you can give to your job, the job being to run \verb|script.r|. This name does not matter, something like ``Attempt 1 to run an R script on Green" will do.

\verb|-pe mpich 1| allows you to decide how many cores you need for your R script. If you need 4 cores, replace \verb|-pe mpich 1| by \verb|-pe snode 4|. If you want 8 cores, replace \verb|-pe mpich 1| by \verb|-pe snode8 8|.

\verb|-l h_vmem=512M| and \verb|-l mem_free=512M| allow you to specify the amount of RAM needed for your script. Other possible values are 1G, 2G, 6G, ...

\verb|-l h_rt=60| is the time in seconds you believe the R code will need to be executed. It doesn't matter if your script is done or not after this time: it will be terminated no matter what. Maximum value is 4320000 (5 days).

\verb|-M me@gmail.com| is your mail. An email will be sent to you once the R script start to run and also when the related R instance is terminated.

\verb|R CMD BATCH /home/smcs/nuyttend/script.r| should be edited according to where your script is located on the cluster.

Once the \verb|job_submission.sh| file is ready, send it to the cluster using (disconnect from the cluster first):

{\scriptsize
\begin{Verbatim}[frame=single, samepage=true]
scp ./job_submission.sh your_login@green.cism.ucl.ac.be:/home/smcs/nuyttend/
\end{Verbatim}
}

Connect yourself to the frontend and submit the two hereafter lines to ask SGE to read \verb|job_submission.sh|:
\begin{Verbatim}[frame=single, samepage=true]
module load sge
qsub job_submission.sh
\end{Verbatim}

The job will be queued. To get some information about the cluster type

\begin{Verbatim}[frame=single, samepage=true]
qload
\end{Verbatim}

To get an idea of how many other jobs there are, type

\begin{Verbatim}[frame=single, samepage=true]
qstat -u "*"
\end{Verbatim}

Once the calculations are done, download back the results like it was done using Hmem.
\subsection{Where is my cluster?}

It is very likely you have a cluster near you (most universities around the world have one). Find out who are the people managing this cluster. Contact them and ask them how to register. Also ask them if they have some examples on how to use the cluster or even better: if they can show a live example on how to use it. Good questions to ask are:

\begin{itemize}[noitemsep, nolistsep] 
\item who can use the cluster? How can I register?
\item is R installed on the cluster?
\item is there a key needed to connect to the cluster? Where can I get it?
\item can I connect from home?
\item what kind of job manager do you use?
\item do you have an example of job submission to the job manager?
\end{itemize}

\section*{Acknowledgments}

Special thanks to Alain Guillet (SMCS, UCL) for allowing me to take over part of his work on the matter a few years ago. The work of Alain was the seed of this paper.

Also many thanks to Damien Fran\c{c}ois from the ``Institut de calcul intensif et de stockage de masse" (UCL) for his help and his work in managing Hmem, Green and many other clusters.

Many thanks to Christian Ritter (UCL) for his support and positive feedback.

Finally, the author of this paper wishes to thank all the researchers working with him at the ``Institut de Statistique, Biostatistique et Sciences Actuarielles'' (UCL) for their many R scripts that are now interesting case studies.

\bibliographystyle{plainnat}
\bibliography{how_to_speed_up_R_code}

\end{document}